# A theoretical framework for dynamic Anticrack and Supershear propagation in snow slab avalanches

**Authors**

Marin SIRON[1,2], Bertil TROTTET[1], Johan GAUME[3,4,5]

[1] School of Architecture, Civil and Environmental Engineering, Swiss Federal Institute of Technology, Lausanne, Switzerland
[2] Ecole Normale Supérieure de Lyon, Université Claude Bernard, Lyon, France
[3] Institute for Geotechnical Engineering, ETH Zurich, Zurich, Switzerland
[4] WSL Institute for Snow and Avalanche Research SLF, Davos Dorf, Switzerland
[5] Climate Change, Extremes, and Natural Hazards in Alpine Regions Research Center CERC, Davos Dorf, Switzerland

## Abstract

Snow slab avalanches release after the failure and collapse of a weak layer buried below a cohesive snow slab. The initial failure is induced by local overloading of the slab, such that the passage of a skier. This results in the propagation of a subsidence, known as a *collapse wave* or *anticrack*. The slab may eventually break and detach from the rest of the snowpack and start to slide down, provided that the slope is steep enough to enable gravity to overcome the friction at the interface between the slab and the failed weak layer.

The approach to anticracks so far has mostly focused on (i) static configurations for the bending slab while the weak layer collapses, thereby leading to analytical conditions for the onset of an anticrack because of the metastability of the snowpack, and (ii) the observation of anticrack propagation as a result of numerical simulation methods (DEM, FEM, MPM) and field experiments (PST). The only theoretical framework to date, based on a simple modelling of the bending of the slab during the weak layer collapse, led to the well-known Heierli (2005) model which suggested an explicit solution for the propagation speed in steady state. It, however, could not account for the weak layer properties, was not mathematically bounded for certain values of the physical constants involved, and could not explain the newly uncovered "*supershear*" transition for steep slopes.

In this paper, a new model for the stationary propagation of anticracks is set up, so as to account for the *anticrack* speed regime on the one hand, and the *supershear* regime on the other hand, the existence of which has been recently revealed and ascertained by numerical simulations. The results presented here seem consistent with most of the available data, and highlight the role that the compaction of the weak layer can play in reducing the anticrack speed. On the contrary, by storing energy upon failure and suddenly releasing it at the crack tip, the weak layer elasticity could help justify the higher speeds sometimes observed in both regimes. Finally, a more accurate model is proposed, based on the modelling of both the slab and the weak layer as Timoshenko beams; although its complexity prevents us from solving it analytically, it provides enlightening insights into the mechanical processes at work at the interface between both layers, from a strength-of-materials perspective.

This analysis is a first step towards a better understanding of the underlying mechanisms of propagation of cracks in slab avalanches, and towards more accurate avalanche size and occurrence predictions.

# Introduction

Contrary to *loose snow avalanches* which typically arise in homogeneous snowpacks with very little cohesion, *slab avalanches* only occur in stratified snow, since they require the collapse of a fragile, sparse sub-layer topped by a dense slab, all of which is supported by a compact snow substrate (as depicted in Figure 1) (Ancey, 2006; Schweizer et al., 2016). If the slope is steep enough, the slab is released following the failure and collapse of the weak underlying layer, breaking into smaller pieces which then slide downwards on the substrate. Because of the extent of the released zone, slab avalanches carry much larger volumes of snow away; thus, from a theoretical standpoint, the in-depth study of their release and flow has received increasing attention in the past decades.

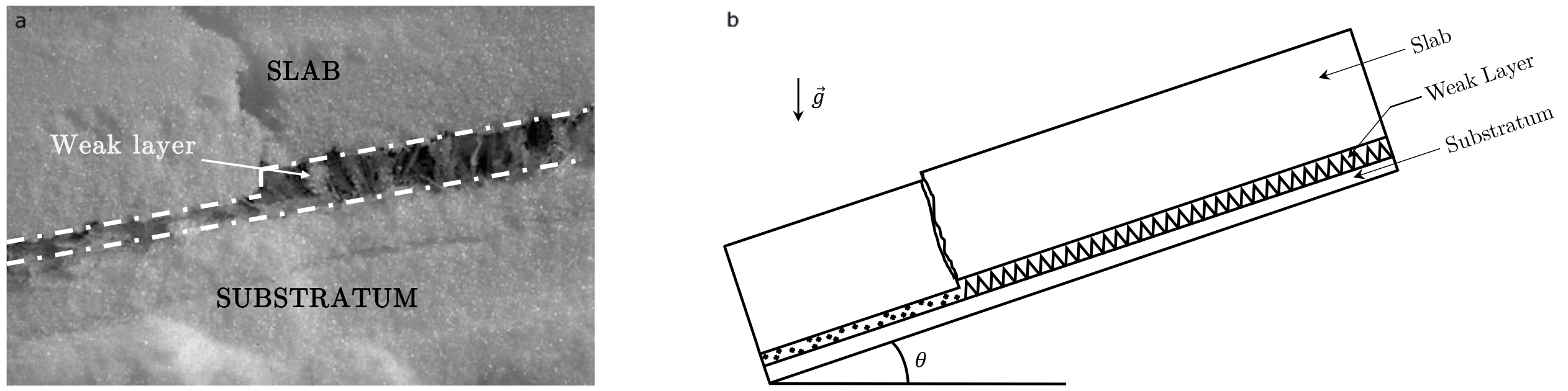

Figure 1 – Stratification of the snowpack conducive to slab avalanche initiation. (**a**) Well-known example of a sample stratified in three layers: the slab overlies a weak layer of partially collapsed surface hoar to the left of a central crack, and intact to the right. On the left-hand side, the partial collapse underlines the metastable aspect of this assembly in the absence of external load (gravity $\vec{g}$ excluded). The dotted lines delineate the layers (low compressible but deformable slab, compressible and deformable weak layer, substratum). © *B. Jamieson (ASARC)* (**b**) Schematic of the situation.

The triggering mechanism of slab avalanches is now well documented. When a sufficient concentrated load is applied, e.g. by a skier or in the wake of a progressive snowfall, the weak layer breaks and locally collapses (Schweizer et al., 2003). It leads to a reduction in its volume - a so-called “volumetric collapse” - whose dynamics itself is still a topic of extensive research, due to the complexity of the constitutive law of snow (Barraclough et al., 2017; Blatny et al., 2022). This compaction of the weak layer causes the overlying slab to bend, which in turn leads to a stress concentration at the edge of the still-intact portion of the weak layer. If this stress is high enough, the edges of the latter part of the weak layer also fail and collapse. This entails the widening of the depression and, as a chain reaction, leads to the self-sustained propagation of the subsidence on both sides of the initial depression (Figure 2).

In terms of fracture mechanics, the propagation of this "collapse wave" manifests an obvious analogy with the widening of a crack at the slab-weak layer interface, although, contrarily to an opening *mode I*, surface lips press against each other, which tends to interpenetrate the slab and the weak layer under the effect of the load. For this reason, (Fletcher & Pollard, 1981) define the notion of *anticrack*, i.e., of a crack propagating under compression, in a closing mode (-I). Thus, the concepts of fracture mechanics are applicable to the collapse process in slab avalanches, provided that the analogy is carried out thoroughly (Heierli et al., 2008). In the following, the "collapse wave" will thus be referred to as an "anticrack".

The condition for the onset of a self-sustained propagating anticrack is generally characterised by a *critical crack length* $a_c$, beyond which the propagation is energetically favourable. It depends on intrinsic parameters of the snowpack, such as the mechanical properties of the layers involved, the friction between them, their dimensions, etc. It can be calculated with the Griffith-Irwin criterion (Heierli et al., 2008; Rosendahl & Weißgraeber, 2020b) or with strength-of-materials methods (eq.(9) in Gaume et al. (2017)). In practice, knowledge of the critical crack length in a snowpack is conspicuously crucial for predicting the probability of triggering of an avalanche and its size. The slope angle, compared to the *crack-face* friction angle, also plays a decisive role in the triggering of an avalanche (van Herwijnen & Heierli, 2009) (Gaume et al., 2013). In both cases, the theoretical determination of these instability criteria can be based on static considerations, since the dynamics of the anticrack is only a consequence of the pre-existing instability. For this reason, the main theoretical investigations so far in avalanche science have focused on static collapse configurations (Benedetti et al., 2019; Gaume, Chambon, et al., 2018; Heierli et al., 2008; Rosendahl & Weißgraeber, 2020).

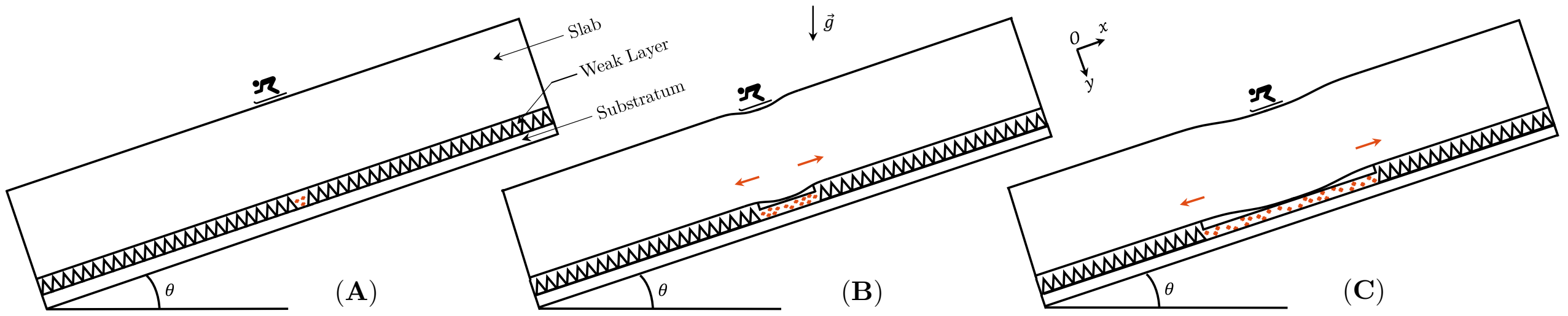

FIGURE 2 – Schematic of the propagation of a *collapse wave* in between a weak sub-surface layer of snow and a slab, at the onset of a snow slab avalanche, rising from birth (**A**) to a self-sustained propagating state (**C**). *Reading*: $\vec{g}$ denotes the acceleration of gravity, and (Oxy) is the frame of reference attached to the slope. The collapsed weak layer is identified by the red crosses. The arrows indicate the direction of propagation. *Note*: the diagram is not to scale. *Adapted from* (Heierli et al., 2008).

Recent contributions nevertheless highlight the interest of studying dynamic configurations to determine the stopping conditions of anticracks (Bergfeld et al., 2022). Besides, with the advent of numerical modelling and the increase in computational power, numerous 'numerical experiments' based on various approaches to avalanche simulation at different scales (DEM: Discrete Element Method; FEM: Finite Element Method; MPM: Material Point Method) have been made possible (Bobillier et al., 2020; Gaume, Gast, et al., 2018; Trottet et al., 2022). They have revealed several particularities specific to the dynamic behaviour of anticracks, in terms of propagation speed especially.

In particular, if a crack propagates while the slope angle is large enough, a transition takes place from a regime of *sub-Rayleigh* propagation velocities, i.e., speeds lower than that of Rayleigh waves ($c_R$) in the slab, to a *supershear* or *intersonic* regime, i.e., speeds between $c_s$ and $c_p$ (shear and dilatational plane wave velocities in the slab, respectively). The mechanism of this transition is known as that of *Burridge-Andrews* (Andrews, 1976; Burridge, 1973). It reflects a jump in the failure mechanism from a mixed mode (compressive -I and in-plane shear II) to a pure shear mode II (Trottet et al., 2022). As of now though, there is little evidence of this *sub-Rayleigh-to-supershear* transition in snow, which has been reported based on avalanche video analysis, and predicted by numerical models. Thus, at present, there is no analytical expression explicitly describing the velocity of the supershear crack as a function of the slope angle.

In the last decade, several studies have attempted to fill this gap:

- On the one hand, Heierli (2005, 2008) has laid the foundations for the dynamics of the slab in free fall, i.e. without taking into account the collapse of the weak layer that causes the slab movement.
  Starting from a simplistic model of a rigid plate in bending, Heierli (2005) asymptotically expresses the anticrack as the self-sustained propagation of a quasi-soliton; he proposes an explicit expression of its propagation velocity as a function of the slab parameters only, since the weak layer is not modelled. In 2021, Bobillier et al. extracted from their numerical experiments an "empirical" correction to this model, by adding terms related to the mechanical behaviour of the weak layer. The additive form of this correction, although consistent with experiments, has no physical foundation.
  In 2008, Heierli added complexity to the previous model by assimilating the slab to a Timoshenko beam, for which bending shear forces are added to those already taken into account. Based on an energetic approach, he derives an open system of differential equations, which couples the propagation velocity to the length of the anticrack. While the approach seems promising, it only provides an analytical expression for the velocity through costly approximations and, again, the weak layer is not accounted for.

- On the other hand, Rosendahl and Weißgraeber (2020a) sought to determine the static form of the slab in bending, by modelling both the slab (as a Timoshenko beam) and the weak layer (as a Winkler support, i.e. an elastic support in shear and compression). Their model led them to detailed criteria for the progression of the anticrack (Rosendahl & Weißgraeber, 2020b). Following energetic considerations, Heierli (2008) had also deduced an expression for the potential energy of the anticrack as a function of its distance from the origin, without attempting to introduce it into a dynamic model.

To our knowledge, however, there exists no theoretical model that simultaneously takes into account the dynamics of the slab and the contribution of the weak layer to theoretically find the collapse wave speed.

# Outline

This study develops an analytical framework in order to characterize the dynamics of mechanical collapse waves (*anticracks*) at the interface between a weak sub-surface layer of snow and an overlying slab, during the onset of snow slab avalanches. In Section I, we assume that an anticrack has been initiated up to the critical length $a_c$, beyond which propagation is self-sustained. Having passed the transient regime, the equations of motion are simplified by assuming the existence of a steady state, in order to find the asymptotic velocities and the parameters on which they depend. Section I details this methodology and its limits as thoroughly as possible. Section II establishes a general formula for the speed of the anticrack in the sub-Rayleigh configuration, and examines its behaviour for different sets of boundary conditions, drawing conclusions on the ones which are physically and mathematically acceptable. It further contains a simple model for the crack in the supershear regime. Section III discusses the previous expressions in light of newly available experimental and numerical data. It also expands on a more accurate modelling of the weak layer, which could account for the fracture process at the front of the anticrack, although its complexity is such that it does not seem feasible to further develop it by analytical means.

# I - Methods

This section presents the methods applied to find the asymptotic propagation speed of the collapse wave in its different configurations, i.e., *sub-Rayleigh* and *supershear*. Obtaining the speed requires to solve simultaneously the equations of motion of the slab, of the weak layer and of the coupling of their dynamics through the boundary conditions. It cannot be achieved simply, thus we decide to simplify the problem by resolving the dynamics of the slab only, and by considering the weak layer as an external resistance applied on the slab.

## A. Finding the equations of motion

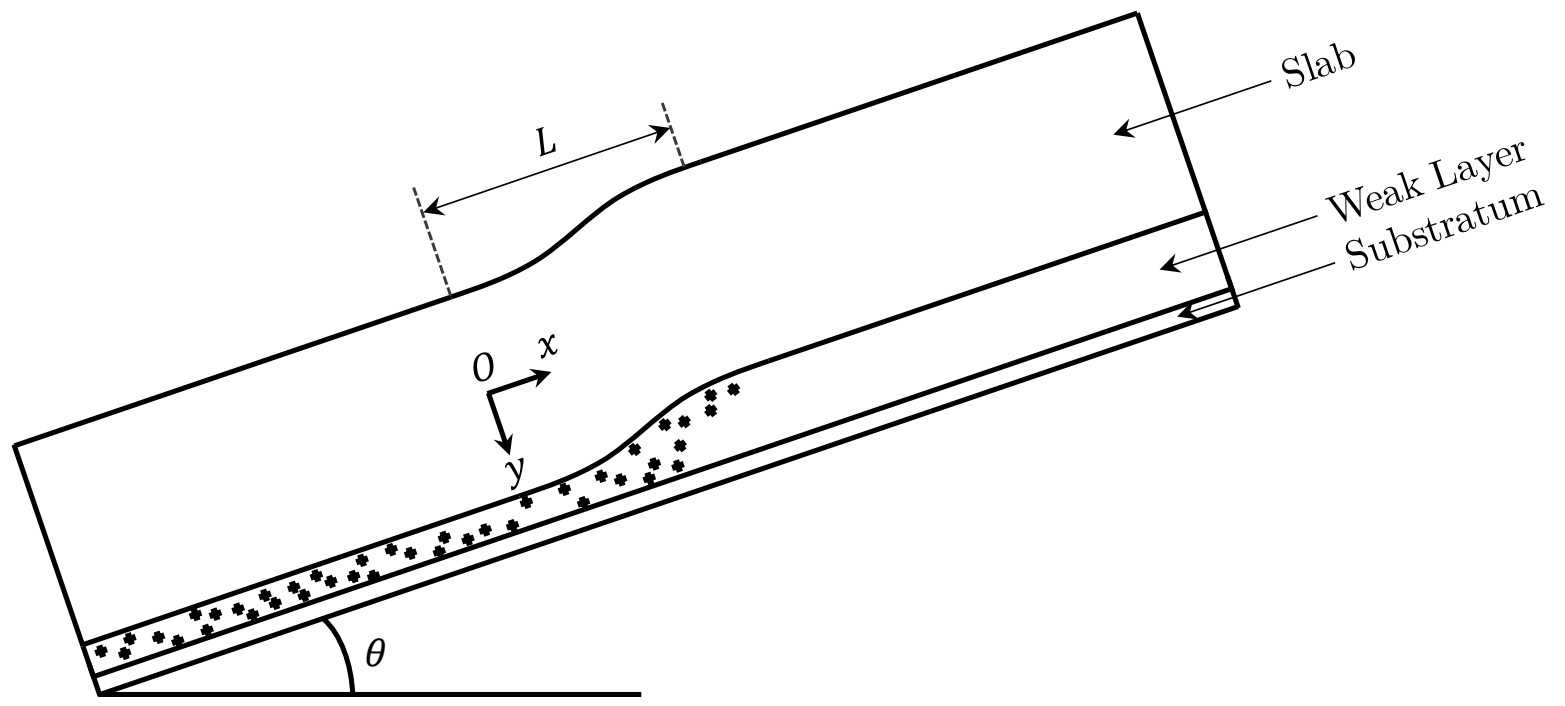

Figure 3 – Schematic of the collapse wave (anticrack) in a two-dimensional approach, valid only at large distances from the origin. The wave travels up the $(Ox)$ axis on a slope inclined at an angle $\theta$ to the ground. The collapse takes place over a distance $L$ such that the origin $O$ of the $(Ox)$ axis can be taken at the point of maximum compaction of the weak layer, and the point $L$ corresponds to the first fracture of the still intact zone of the weak layer. The crosses mark the degraded zone of the weak layer (fracturing, compaction).

At first, a mechanical model of the slab has to be chosen. Following the work of Heierli (2005 and 2008) and of Rosendahl and Weißgraeber (2020a), the slab can be modelled either as a plate or as a beam, depending on the spatial extensions considered relevant.

For the sake of simplicity, in the following, we consider the slope as an infinite plane inclined by an angle $\theta$ with respect to the ground (see Figure 3). Also, at large distances from the origin of the crack, the wave fronts become planes perpendicular to the $(Ox)$ axis of propagation; by defining the axis reference frame as the $(Oxy)$ plane for the collapse wave, it seems reasonable to assume invariance of the deformations in the $(Oz)$ direction. As the problem becomes two-dimensional, the notions of plates and beams become identical. Nevertheless, it remains to choose the orientation of the propagation axis $(Ox)$ with respect to the axis of the slope. In reality, the crack propagates in all directions from the origin of the failure, following a mixed mode (-I, II: in-plane shear, and III: out-of-plane shear) with a strong dependence on the terrain features crossed at each moment. The propagation of the anticrack is therefore different depending on whether it propagates mainly parallel or perpendicular to the slope, because the gravity components projected in the plane of the slope are distinct in these cases. In the following, the crack will be assumed to travel parallel to the slope (upwards or downwards) without loss of generality, given that any other orientation will be recoverable from our results by modifying gravity accordingly.

Modelling the slab as a so-called "Timoshenko beam" has the advantage of taking into consideration the internal shear forces in the slab; it differs from the Euler-Bernoulli beam by the absence of the traditional assumption of perpendicularity of the plane sections of the beam (or *cross sections*) to the beam generatrix (namely, the axis joining the middle of the cross sections). More comprehensive models exist (e.g., Levinson beams), yet, with the aim of facilitating the comparison of our results with the literature, we will adopt a Timoshenko beam model in the following. As for any beam model, it imposes a decoupling between the vertical and longitudinal dependencies of the displacement fields. Its kinematics assumes the following relationships for the displacement fields $u_x(x,y,t)$ and $u_y(x,y,t)$ at any point on the beam:

$$u_x(x,y,t) = u(x,t) - y \cdot \psi(x,t)$$
$$u_y(x,y,t) = v(x,t)$$

where $u(x,t)$ et $v(x,t)$ are the displacement fields in the middle of the beam sections. In steady state, the collapse wave velocity can be determined from the evolution of any point of the beam. For convenience, the generatrix of the beam is chosen, through which the velocity can be determined directly from the equations on $u(x,t)$, $v(x,t)$ and $\psi(x,t)$. Figure 4 summarises these assumptions in a synthetic view.

In order to obtain the equations of motion of the slab, we decide to use Hamilton's Principle (or variational principle) defined from the action $\mathcal{S}$:

$$\mathcal{S}(u,v,\psi,t) = \int_t \mathcal{L}(u(x),v(x),\psi(x),t)\ dt$$

where $\mathcal{L} = \mathcal{E}_{kinetic} - \mathcal{E}_{potential}$ denotes the Lagrangian of the system and $u, v, \psi$ are the displacement fields of the Timoshenko beam explained in Figure 4. The principle is stated in Supplementary Material n°1.

Once this model is defined, the kinetic and cohesive energies of the beam are to be determined. Due to the scale factor of the problem (the height of the slab is very large compared to the height of the weak layer), the deformations during collapse are small, and the associated forces can be taken in their linear limit. Thus,

- Kinetic energy: $b \cdot h \cdot \int_0^L \left\{\frac{1}{2}\rho(\frac{\partial u}{\partial t})^2 + \frac{1}{2}\rho(\frac{\partial v}{\partial t})^2 + \frac{1}{2}\rho I\left(\frac{\partial \psi}{\partial t}\right)^2\right\} dx$
- Tensile-compressive potential energy: $b \cdot h \cdot \int_0^L \{\frac{1}{2}E(\frac{\partial u}{\partial x})^2\} dx$
- Flexural potential energy: $b \cdot h \cdot \int_0^L \left\{\frac{1}{2}EI\left(\frac{\partial \psi}{\partial x}\right)^2\right\} dx$
- Potential energy for shear strains: $b \cdot h \cdot \int_0^L \{\frac{1}{2}\kappa G(\frac{\partial v}{\partial x} - \psi)^2\} dx$. This secondary effect of flexion is specific to the Timoshenko beam theory.

with the following notations:

- $[0, L]$ defines the bending section (the part where the weak layer collapses);
- $A = b \cdot h$ is the surface of a cross section of the beam, with: $h$, the height of the beam/slab; $b$, the width of the beam.
- $I = \frac{h^2}{12}$, second moment of area of the undeformed beam (rotational moment with respect to the (Oy) axis), normalised by the surface $A = b \cdot h$;
- $E = \frac{E'}{1-\nu^2}$, plane strain Young's modulus of the beam (linked to Young's modulus $E'$); $G$, plane strain shear modulus of the beam ($G = \frac{E}{2(1+\nu)} = \frac{E'}{2(1+\nu)(1-\nu^2)}$);
- $\kappa$, called the *Timoshenko correction factor*, is equal to 5/6 for a rectangular beam.

Since the beam model allows us to neglect the dependency of the crack on the $(Oz)$ dimension, *for convenience we will choose* $b = 1$ *throughout the rest of the paper*, without loss of generality.

Finally, one should add to these energies those of the external forces applied to the beam. This obviously includes gravity, but also contributions from the underlying weak layer, in particular the fracture energy required to fracture the weak layer during crack propagation. The modelling of this contribution is the subject of the following section.

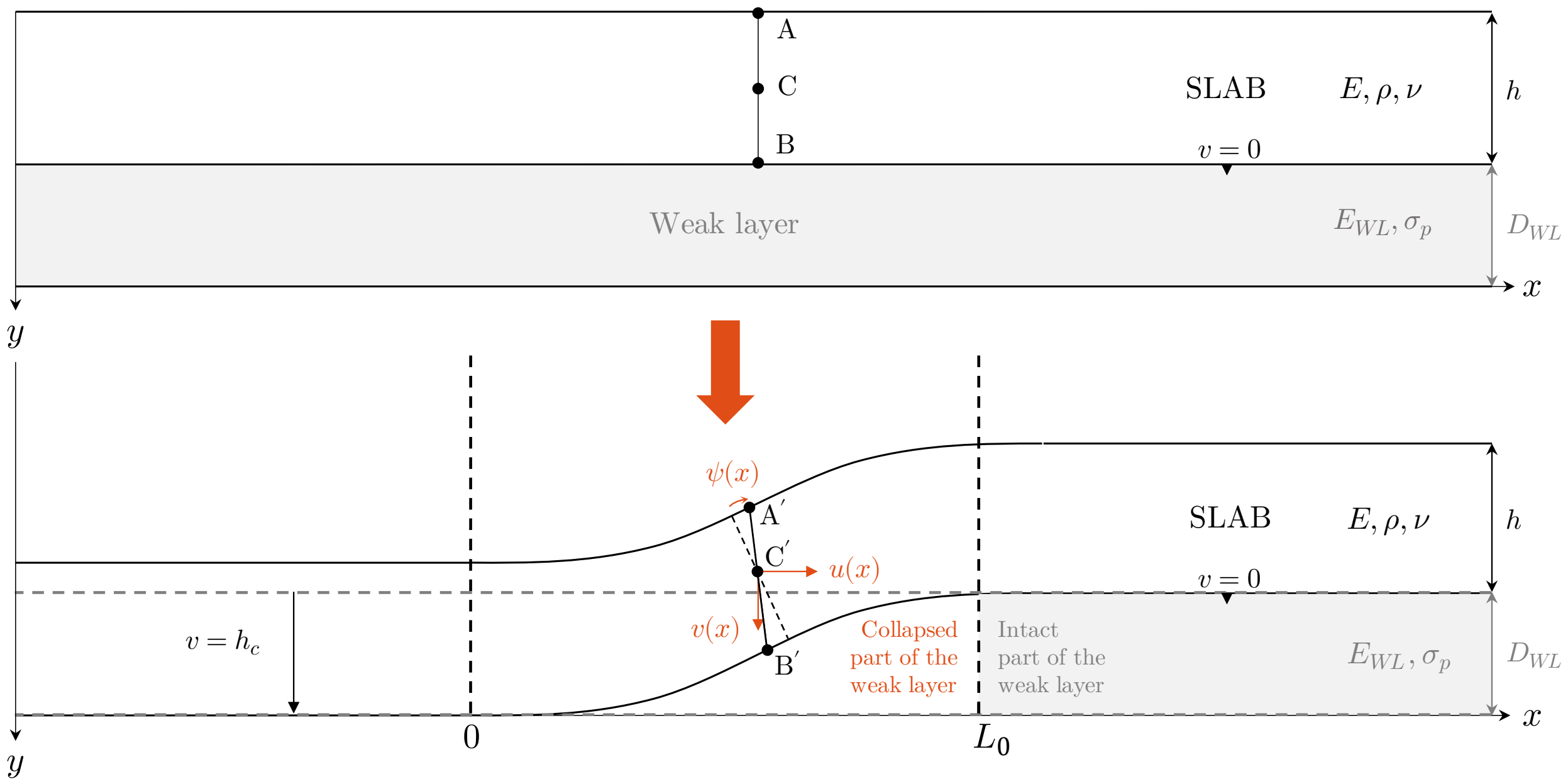

FIGURE 4 – Diagram of slab deformation in a Timoshenko beam model for a crack of length $L_0$ at time $t$. (**Top**) Intact stratified snowpack before crack initiation. (**Bottom**) Slab deflection after the collapse of the weak layer over a crack distance $L_0(t)$ on either side of the deflection. *Reading*: any plane section AB deformed at A'B' remains plane, but tilts by an angle $\psi$, with translation of its centre C by $u.\vec{x}$ and $v.\vec{y}$ to C'. The displacement fields $u(x,t)$, $v(x,t)$ and $\psi(x,t)$ depend only, in this configuration, on the longitudinal coordinate $x$ and time $t$. *Note*: the diagram is not to scale.

## B. Modelling the weak layer

The objective of this section is to establish an analytical expression for the asymptotic propagation velocities of the anticrack. To do so, the perturbation is followed at a great distance from its origin, and is assumed to have reached a stationary regime of propagation along increasing $x$. In a Galilean reference frame attached to the ground, each point of the perturbation is in uniform translation along the $(Ox)$ axis at the speed $c$. In the mobile reference frame of axis $(0x')$ linked to the disturbance, the beam appears as static, and the disturbance occupies a fixed section $]0, L[$ in this reference frame, where $L = L_0 - ct$. Figure 5 shows the situation in this moving frame of reference.

In the end, the problem is a combination of the solutions obtained on the three sections considered separately.

### 1. Complete formulation of the problem

Without taking into account the work of the fracturing forces applied by the weak layer on the slab, the evolution of the displacement fields of the slab takes the following form (Heierli, 2008). On the bending section $[0, L_0(t)]$ where the collapse happens:

- *Action functional*:

$$S(u,v,t) = h\int_0^t \int_0^L \left\{ \frac{1}{2}\rho\left(\frac{\partial u}{\partial t}\right)^2 + \frac{1}{2}\rho\left(\frac{\partial v}{\partial t}\right)^2 + \frac{1}{2}\rho I\left(\frac{\partial \psi}{\partial t}\right)^2 - \frac{1}{2}E\left(\frac{\partial u}{\partial x}\right)^2 - \frac{1}{2}E\left(\frac{\partial v}{\partial x}\right)^2 - \frac{1}{2}\kappa G\left(\frac{\partial v}{\partial x} - \psi\right)^2 + \frac{\tau}{h}u - \frac{\sigma}{h}v \right\} dx\, dt$$

  with $\sigma = -\rho g h \cdot cos(\theta)$ the (negative) compressive stress due to the beam's own weight, and $\tau = \rho g h \cdot sin(\theta)$ the shear stress due to the weight of the beam; one recalls that $b$ has been omitted by being taken equal to 1.

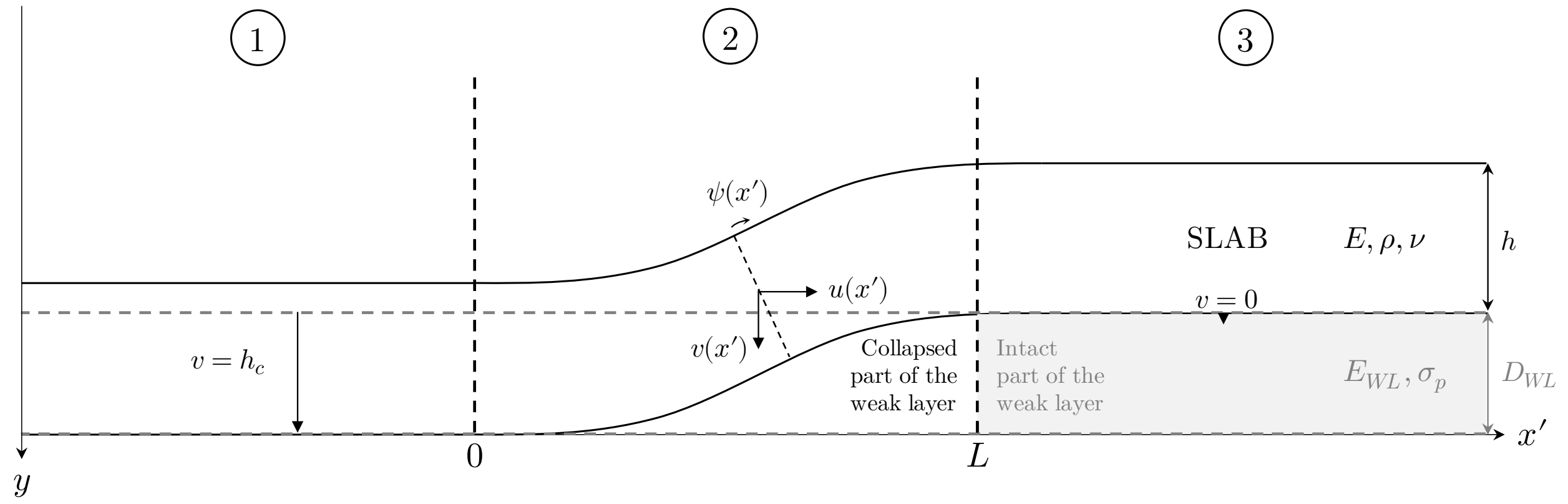

FIGURE 5 – Schematic of the steady state disturbance in the moving reference frame attached to the collapse wave moving from left to right at constant speed. The slab (a Timoshenko beam) can be divided into three sections. From right to left: (*Section 3*) From $+\infty$ to $x' = L$, the beam is supported by the still intact weak layer. At $x' = L$, the weak layer is suddenly fractured due to stress accumulation. (*Section 2*) From $x' = L$ to $x' = 0$, the fractured weak layer is collapsing under the weight of the overlying slab (*volumetric collapse*). The slab bends accordingly until $x' = 0$, defined as the point of tangency where it rests on the substrate after having completely compacted the weak layer. (*Section 1*) From $x' = 0$ to $-\infty$, the disturbance has passed and the weak layer has been completely compacted. The beam rests uniformly on the substrate. *Note*: the diagram is not to scale.

- *Continuity conditions for the connection to the other sections*:
  *At time t* (time dependency of the variables is therefore omitted in the following, for easier writing),

  - Continuity of the displacement fields at $x = 0$ and $x = L_0$ :

$$\begin{aligned} u(0^-) &= u(0^+) \\ v(0^-) &= v(0^+) \\ \psi(0^-) &= \psi(0^+) \end{aligned} \qquad \begin{aligned} u(L_0^-) &= u(L_0^+) \\ v(L_0^-) &= v(L_0^+) \\ \psi(L_0^-) &= \psi(L_0^+) \end{aligned}$$

  - Continuity of the constraints at $x = 0$ and $x = L_0$ :

  Using the above-mentioned kinematics, the forces of the Timoshenko beam for low deformations are defined by:

  - The bending moment $\mathcal{M} = -EAI\psi'(x,t)$
  - The transverse shear force $\mathcal{Q} = \kappa GA \cdot \big(v'(x,t) - \psi(x,t)\big)$
  - The longitudinal elastic force $\mathcal{N} = EAu'(x,t)$

  Their continuity at the edges implies that of $v' - \psi$, $u'$, $\psi'$ and, by making use of the continuity of the displacement fields, that of $v', u', \psi'$ as well:

$$\begin{aligned} u'(0^-) &= u'(0^+) \\ v'(0^-) &= v'(0^+) \\ \psi'(0^-) &= \psi'(0^+) \end{aligned} \qquad \begin{aligned} u'(L_0^-) &= u'(L_0^+) \\ v'(L_0^-) &= v'(L_0^+) \\ \psi'(L_0^-) &= \psi'(L_0^+) \end{aligned}$$

- *Boundary conditions for stresses*: in addition to the guaranteed continuity of constraints at the edges, the value of these constraints can be specified using the variational principle. Free edges, for instance, require vanishing moments and internal forces, leading to $v' = \psi' = u' = 0$ in such cases.

The previous equations and conditions can then be simplified by using the moving frame of reference, which results in the removal of the time dependence and in the substitution of $L_0(t)$ by $L$.

## 2. Adding a fracture force

The modelling of the fracturing process, that must be incorporated into the equations of motion, is the main novelty of this study. As the equations are derived from an energy reasoning, we seek a local expression for the energy dissipated during snow fracturing as the anticrack propagates.

The physical processes involved during brittle layer failure have recently been summarised by Bergfeld et al. (2022), whose comprehensive explanatory diagram is adapted in Figure 6. Note that this description is only valid in the sub-Rayleigh propagation regime, where the anticrack is essentially reduced to a transverse disturbance.

As a granular medium, the snow making up the weak layer consists of *load* (or *force*) *chains*, i.e. preferential paths of connected grains that channel stress in the layer under load. Bergfeld et al. (2022) highlight that the fracturing of the weak layer at the *crack tip* (or crack front) is microscopically linked to the rupture of these load chains when the disturbance passes. The arrival of the anticrack on an intact zone of the weak layer generates the first fracturing of all the pre-existing load chains: this step, which occurs at the front of the disturbance, is the first to be energy-consuming. At the end of this initial fracture, in the sub-Rayleigh regime, the debris of the weak layer are compacted by the rest of the disturbance. However, snow crystals in these debris reorganise very frequently and create new bonds between each other, which must be broken continuously to compact them: this requires an ever-renewed energy during compaction, to develop secondary cracks in these new bonds. Thus, the compaction of the weak layer takes place progressively, until it stops when the packing of the weak layer is too strong to ensure further compaction. Based on this observation, Bergfeld et al. (2022) identify two different fracturing zones, associated with two distinct sources of energy consumption:

- The *dissipation of dynamic fracture*: it accounts for the initial energy supplied to the front of the anticrack to break the initial force chains, break the bonds at the head of the crack tip, and generate local plastic deformations.
- The *dissipation of compaction*: downstream of the anticrack front, the weak layer is progressively compacted, calling for the provision of adequate energy, microscopically linked to the friction and bond-breaking during this compaction.

These two energies must now be expressed. Two different visions can be used to achieve it.

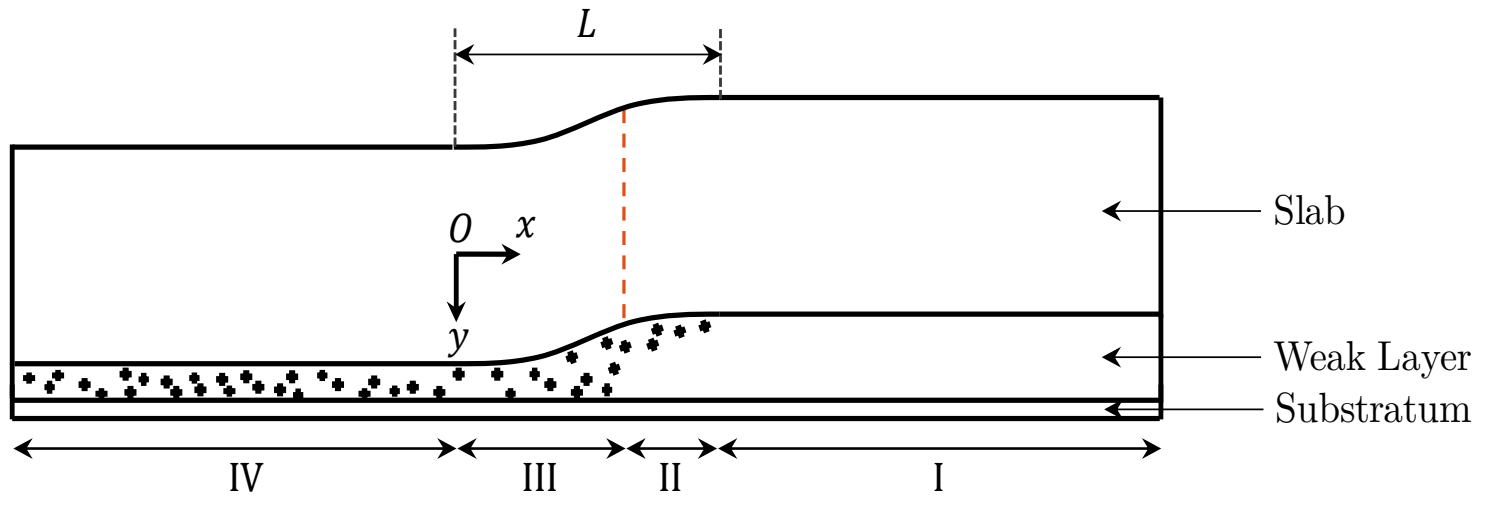

Figure 6 - Schematic representation of a propagating *anticrack* on flat terrain far from its origin. *Reading*: "The crack tip propagates from left to right. In **region I**, weak-layer bonds are not yet fractured, while in **region IV** all weak-layer bonds are broken. **Region II** is the fracture process zone, extending from the first bond fractures to the crack tip. In **region III**, the slab further subsides causing the weak-layer structure to fracture multiple times before closer packing of the weak layer is achieved and the slab comes to rest again." The orange dotted line refers to the position of the crack tip, while black marks denote broken bonds; $L$ is the touch-down distance. © *Adapted from* (Bergfeld et al., 2022), *with its original legend.*

### a) *Through fracture mechanics*

One interest in having expressed the collapse of the weak layer as an anticrack (a 'compressive fracture', whose lips close rather than open, as opposed to Mode I) is the analogy it offers with the tools of fault mechanics. Indeed, it is possible to use failure criteria (e.g., the Griffith criterion) so as to specify the work required to "open" an anticrack over a length $dx'$ in the weak layer of arbitrary thickness $b$ :

$$W_f^{dyn} = w_f \cdot b \cdot dx'$$

By analogy, this clearly embodies the *dissipation of dynamic fracture.* In this expression, $w_f$ therefore refers to the specific fracture energy (per unit of crack area, in $J.m^{-2}$) to be provided to remove bonds and create a $b \cdot dx'$ surface of anticrack in the weak layer. As pertains to the *dissipation of compaction*, it can be expressed in the same way:

$$W_{comp} = w_{comp} \cdot b \cdot dx'$$

Formulated in this way, however, these works are not convenient to use in our problem, as they do not explicitly depend on any displacement field of the disturbance $(u, v, \psi)$, although it is clear that the energy dissipated by the compaction of the weak layer increases with the collapse height of the latter. In other words, since the compaction occurs in volume and not in surface, a volumetric compaction energy is more suitable for the situation. Following the assumption made by Bergfeld et al. (2022), we can then write:

$$\delta W_{comp} = w_{comp}^{vol} \cdot b \cdot v(x) \cdot dx' \qquad \forall x' \leq L \tag{1}$$

Moreover, it has been shown (Rosendahl & Weißgraeber, 2020a) that the fracture process at $x' = L$ can be split into two contributions, since it occurs in *mixed-mode* (modes -I and II). Thus,

$$\delta W_f^{dyn} = w_{f,I} \cdot b \cdot dx' + w_{f,II} \cdot b \cdot dx' \quad at\ x' = L \tag{2}$$

As *Region II* of Figure 6 is narrow, it results in two quasi-point forces associated with these energies at $x' = L$ and given by:

$$\overrightarrow{f_I} = -\, w_{f,I} \cdot b\, \overrightarrow{e_{x'}} \qquad\qquad \overrightarrow{f_{II}} = -\, w_{f,II} \cdot b\, \overrightarrow{e_y}$$

where $w_{f,I}$ and $w_{f,II}$ are specific energies of dynamic fracture $(J.m^{-2})$ for modes -I and II, respectively. Bergfeld et al. (2022) finally underline, based on PST experiments, that the specific work of compaction is typically *thirty times higher* than the specific dissipation of dynamic fracture: compaction thus seems to be the dominant dissipative process. For this reason, it will be considered that $\forall x' \leq L$,

$$\delta W_{fracture} \approx \delta W_{comp} = w_{comp}^{vol} \cdot b \cdot v(x) \cdot dx' \tag{3}$$

**In summary:** This model, based on fracture mechanics and relevant for the *sub-Rayleigh* regime, gives rise to the results presented in Section II-A.

### b) *Through the strength of materials*

The rheology of snow is particularly complex, and can be caught by an elastoplastic (Blatny et al., 2022; Gaume, Gast, et al., 2018) or even elastoviscoplastic behaviour (Cresseri et al., 2009). However, uni-axial compression and tensile tests of the weak layer can be used to extract a simplified constitutive law (see Figure 7) (Grégoire Bobillier, 2022).

In the sub-Rayleigh regime, where the transverse deformations of the beam prevail, the fracturing of the weak layer can be considered to take place in compression mainly, and to be modelled as a brittle failure (green dotted curve in Figure 7): this alternative model is derived in Supplementary Material n°4 and briefly analysed in Section III-A.3.

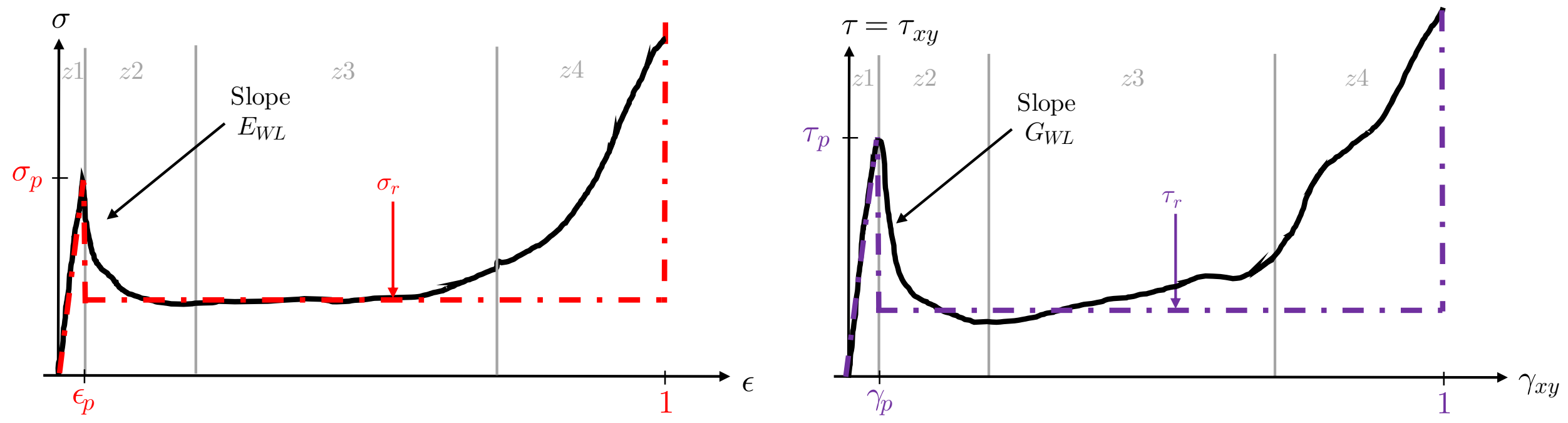

FIGURE 7 – (**Left**) Behaviour of the weak layer during a compression test under controlled load. The blue curve corresponds to the normal stress during the four phases of the collapse (elastic phase in zone $z.1$, followed by failure with softening in $z.2$; then brittle compaction in $z.3$ and densification in $z.4$). The red dotted curve corresponds to the model adopted in the *sub-Rayleigh* regime. (**Right**) Behaviour of the weak layer during a load-controlled shear test. The purple dotted curve now corresponds to the model adopted in the *supershear* regime, when the anticrack is in shear. *Adapted from the DEM simulations of* Bobillier (2022).

In the supershear regime, where, conversely, longitudinal deformations are preponderant, we consider that the fracturing of the weak layer takes place in shear only, and that it can be modelled as a brittle failure too: this last model is developed in Supplementary Material n°3.

### c) *Combining both approaches*

The two previous views are compatible as long as they are linked through the constitutive laws used in strength of materials. An example is given in Figure 8 for a snowpack loaded in compression, which allows the volumetric energies to be expressed as areas under the constitutive law. As a first approximation, neglecting the plastic energy input by strain softening (area $z2$ in Figure 7) leads to a brittle fracture in compression, thus

$$w_{f,mode\ -I}^{vol} = \frac{1}{2}\sigma_p \epsilon_p = \frac{\sigma_p^2}{2E_{WL}} \tag{4}$$

where $E_{WL}$ is the Young's modulus of the weak layer, $\sigma_p$ its peak stress at break (equal to its yield strength), $\epsilon_p$ its peak strain at break and $D_{WL}$ its height. From this point of view, the work to be done to fracture a section of length $dx'$ through a thickness $D_{WL}$ is given by the elastic energy:

$$W_{f,dyn} = \frac{1}{2}\sigma_p \epsilon_p \cdot b \cdot D_{WL} \cdot dx' \tag{5}$$

whence it can be derived the specific energy of dynamic fracture:

$$w_{f,I} = \frac{1}{2}\frac{\sigma_p^2}{E_{WL}} D_{WL} \tag{6}$$

Obviously, if the fracture occurs in shear, the expression for the volume fracture energy (which has to be injected into Equation (2)) is modified according to:

$$w_{f,mode\ II}^{vol} = \frac{1}{2}\tau_p \gamma_p = \frac{1}{2}\frac{\tau_p^2}{G_{WL}} D_{WL} \tag{7}$$

where $G_{WL}$ is the shear modulus of the weak layer, $\tau_p$ its peak stress at break (equal to its shear yield strength), $\gamma_p$ its peak shear strain at break and $D_{WL}$ its height.

Finally, for a mixed-mode anticrack, it can again be rewritten: $w_f^{vol} = w_{f,mode\ -I}^{vol} + w_{f,mode\ II}^{vol}$.

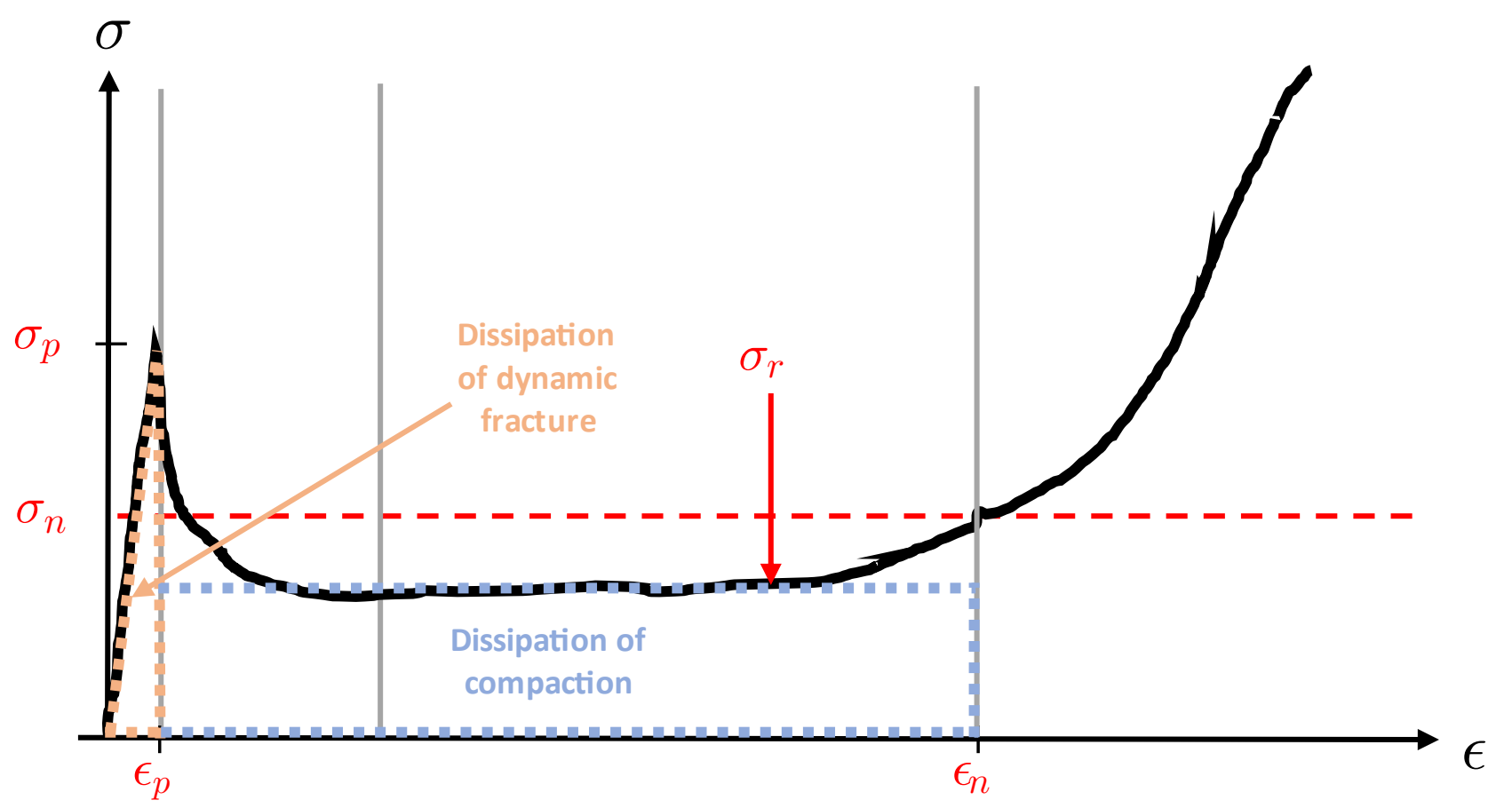

FIGURE 8 – Behaviour of the weak layer during a load-controlled compression test (black line), and identification of the volumetric *dissipation of compaction* and *dissipation of dynamic fracture*. When compaction occurs under realistic conditions (namely, during the collapse of the weak layer caused by the passage of an anticrack), the final stress $\sigma_n$ (final strain $\epsilon_n$, respectively) is determined by the amount of gravitational potential energy available in the slab in order to compact the weak layer. After the collapse, since the slab applies a constant stress equal to its weight when it eventually rests on the compacted weak layer, it can be inferred that $\sigma_n = \rho ghcos(\theta)$.

# II - Results

In this section, we present the solutions of the problem solved with constant fracture and compaction energies, in a mixed perspective between fracture mechanics and strength of materials. The *sub-Rayleigh* and *supershear* models are treated in separate sections.

## A. Sub-Rayleigh regime

This first model is based on fracture mechanics at speeds for which transverse strains $v$ prevail over longitudinal strains $u$ ($c < c_s$). As the compressive mode (-I) is predominant, we assume that the speed of the anticrack is driven by that of a flexural wave propagating in the slab, while the latter undergoes only one constant contribution from the weak layer, namely a delocalised *work of compaction* on the entire $[0, L]$ section:

$$\delta W_{comp} = w_{comp}^{vol} \cdot b \cdot v(x) \cdot dx'$$

Now that all the forces involved in the disturbed section are known, the equation of the motion of the bending in the central section can be determined (see details in Section 2 of Supplementary Material n°2):

$$\left(1 - \frac{c^2}{c_s^2}\right)\left(1 - \frac{c^2}{c_p^2}\right) v^{(4)} + \frac{1}{\lambda^2}\frac{c^2}{c_s^2} v^{(2)} = \frac{1}{\lambda^2 h}\Sigma_{eff} \tag{8}$$

where:

- $c_s = \sqrt{\frac{\kappa G}{\rho}}$ denotes the speed of two-dimensional shear S-waves in our Timoshenko slab;
- $c_p = \sqrt{\frac{E}{\rho}}$ denotes the speed of two-dimensional dilatational P-waves in the slab;
- $\lambda = \sqrt{\frac{EI}{\kappa G}}$ is a characteristic wavelength of the disturbance;
- $\Sigma_{eff} = \frac{\rho gh \cos(\theta) - w_{comp}^{vol}}{\kappa G}$ is the normalised effective gravity stress.

The problem must be closed by a set of boundary conditions. Six are required to close the dynamic problem, and only five for the static problem (see Supplementary Material n°6 for further details). The choice of their values is crucial, since it completely determines the velocities that can be reached.

## 1. Generic boundary conditions

Six boundary conditions must be brought forth to account for the six unknowns on the central collapsing section (namely, the four integration constants $A, B, \mathcal{C}, D$, the speed $c$ and the touch-down distance $L$). To this end, we recall the conditions used by:

- Heierli (2005) :

$$v(x'=0)=h_c \qquad \frac{dv}{dx'}(x'=L)=0$$
$$v(x'=L)=0 \qquad \frac{d^2v}{dx'^2}(x'=0)=0$$
$$\frac{dv}{dx'}(x'=0)=0 \qquad \frac{d^2v}{dx'^2}(x'=L)=-\frac{g}{c^2}$$

- Rosendahl & Weißgraeber (2020a) : Static conditions at the junction of the sections require continuity of the displacement fields and their derivatives at $x'=0$ and $x'=L$.

In order to provide the most general results possible, we take for granted the first three conditions of Heierli (2005), and leave the last three generic so as to match Rosendahl and Weißgraeber (2020a)'s perspective:

$$v(x'=0)=h_c \qquad \frac{1}{h}\frac{dv}{dx'}(x'=L)=V'_L$$
$$v(x'=L)=0 \qquad \frac{d\psi}{dx'}(x'=0)=\psi'_0$$
$$\frac{dv}{dx'}(x'=0)=0 \qquad \frac{d\psi}{dx'}(x'=L)=\psi'_L$$

Obtaining the velocity explicitly is lengthy and the derivations are given in Section 5 of Supplementary Material n°2. At the end of the calculations, speed appears as a function of both the system parameters and the boundary conditions, and takes the following form:

$$\frac{c_s^4}{c^4}\lambda^2\left(1-\frac{c^2}{c_p^2}\right)\frac{\Sigma_{eff}}{h^2}\left[\gamma\tan\left(\frac{\gamma}{2}\right)\left(1-\frac{\psi'_0}{\frac{c_s^2}{c^2}\frac{\Sigma_{eff}}{h}}\right)+\left[\frac{\gamma}{\sin(\gamma)}-1\right]\frac{\psi'_0-\psi'_L}{\frac{c_s^2}{c^2}\frac{\Sigma_{eff}}{h}}-\frac{1}{2}\gamma^2\left(1-\frac{c^2}{c_s^2}\right)\right]=-\frac{h_c}{h} \tag{9}$$

$$\frac{c_s^2}{c^2}\frac{\Sigma_{eff}}{h^2}\left[\gamma\left(1-\frac{c^2}{c_s^2}\right)-2\tan\left(\frac{\gamma}{2}\right)\right]=-\tan\left(\frac{\gamma}{2}\right)\left(\frac{\psi'_0+\psi'_L}{h}\right)+\gamma\frac{\sqrt{1-\frac{c^2}{c_s^2}}}{\sqrt{1-\frac{c^2}{c_p^2}}}\frac{c}{c_s}\frac{1}{\lambda}V'_L \tag{10}$$

$$L=\gamma\lambda\cdot\frac{c_s}{c}\sqrt{\left(1-\frac{c^2}{c_s^2}\right)\left(1-\frac{c^2}{c_p^2}\right)} \tag{11}$$

Equations (9) and (10) consist of coupled equations on the variables $\gamma$ and $c$, where $\gamma$ was defined for convenience and is linked to the touch-down distance $L$ and the anticrack speed $c$. Their decoupling cannot be achieved through analytical means, due to the generic boundary conditions and the equations' non-linearity; yet; their numerical resolution is straightforward and allows to find both the speed $c$ and the touch-down distance $L$ through the determination of $\gamma$.

The next section examines a particular case in which the analytical decoupling is made possible, leading to a generalisation of the well-known solution of Heierli (2005).

## 2. Generalised "Heierli Solution"

To begin with, we consider the boundary conditions (6b) and (6c) of Heierli (2005):

$$v(0) = h_c \qquad v(L) = 0$$
$$\frac{dv}{dx'}(0) = 0 \qquad \frac{dv}{dx'}(L) = 0$$
$$\frac{d^2v}{dx'^2}(0) = 0 \qquad \frac{d^2v}{dx'^2}(L) = \frac{g}{c^2}$$

Injecting them into Equations (9) and (10) results in Equations (12) to (14) below.

$$c = c_s \sqrt[4]{\frac{1}{2}\frac{E}{12(\kappa G)^2}\frac{h}{h_c}\left(\rho g h \cos(\theta) - w_{comp}^{vol}\right)\left(1-\frac{c^2}{c_p^2}\right)\left(1-\frac{c^2}{c_s^2}\right)} \tag{12}$$

$$L = \gamma \cdot \lambda \cdot \frac{c}{c_s}\sqrt{\left(1-\frac{c^2}{c_s^2}\right)\left(1-\frac{c^2}{c_p^2}\right)} \tag{13}$$

$$\gamma = \tan\left(\frac{\gamma}{2}\right) \approx 2.3311 \tag{14}$$

Let us recall the formulas for bending length $L$ and velocity $c$ found by Heierli (2005) (Eq. 7a) and updated in our notation system:

$$c^4 = \frac{g}{2h_c}\frac{Eh^2}{12\rho} \tag{15}$$

$$L^4 = 2.331^4\frac{2h_c}{g}\frac{Eh^2}{12\rho} \tag{16}$$

This set of equations will be referred to as the *Heierli solution* in the rest of the paper, while our equations (12) to (14) will be called the *Generalised Heierli solution*.

Several remarks can be made:

- Equation (12) is a bi-squared fourth-order polynomial in the anticrack speed $c$, which means that it is theoretically possible to express its roots, although it is cumbersome, thus not detailed here for clarity. Only one positive root is expected in between 0 and $c_s$ (as numerically ascertained).

- We recover the *Heierli solution* by considering the low-speed limit ($c \ll c_s$), no compaction process ($w_{comp}^{vol} = 0$) and a flat terrain ($\theta = 0$) in the above set of equations. It should be understood here that the equation provided by Heierli (2005) is only valid for low speeds, although no data on its range of validity was provided by the author. Please note, however, that the touch-down distance $L$ is still different, since our Timoshenko model considers additional internal shear forces that modify the expression of $\lambda$, compared to Heierli's.

- It can be highlighted from Equation (12) that the dissipation of compaction directly competes against gravity, so that it affects speed as though the slab was subjected to an effective gravity (smaller than $g$) during its fall; or equivalently, as though it underwent a slope angle lower than expected. The dissipation of compaction therefore appears as an *additive* term in the formula. As outlined in Section I-B.2.c, linking the *work of compaction* (a relevant quantity for fracture mechanics) and the mechanical properties of the weak layer could contribute to give a physical basis to the additive form of the corrective terms to the *Heierli solution*, which was introduced by Bobillier (2022) for numerical simulations. Note that the dependency on the compaction process is still relatively low for classical values of ~100 $J.m^{-3}$ for the dissipation of compaction (Bergfeld et al., 2022), but significant for higher values, as displayed in Figure 9.

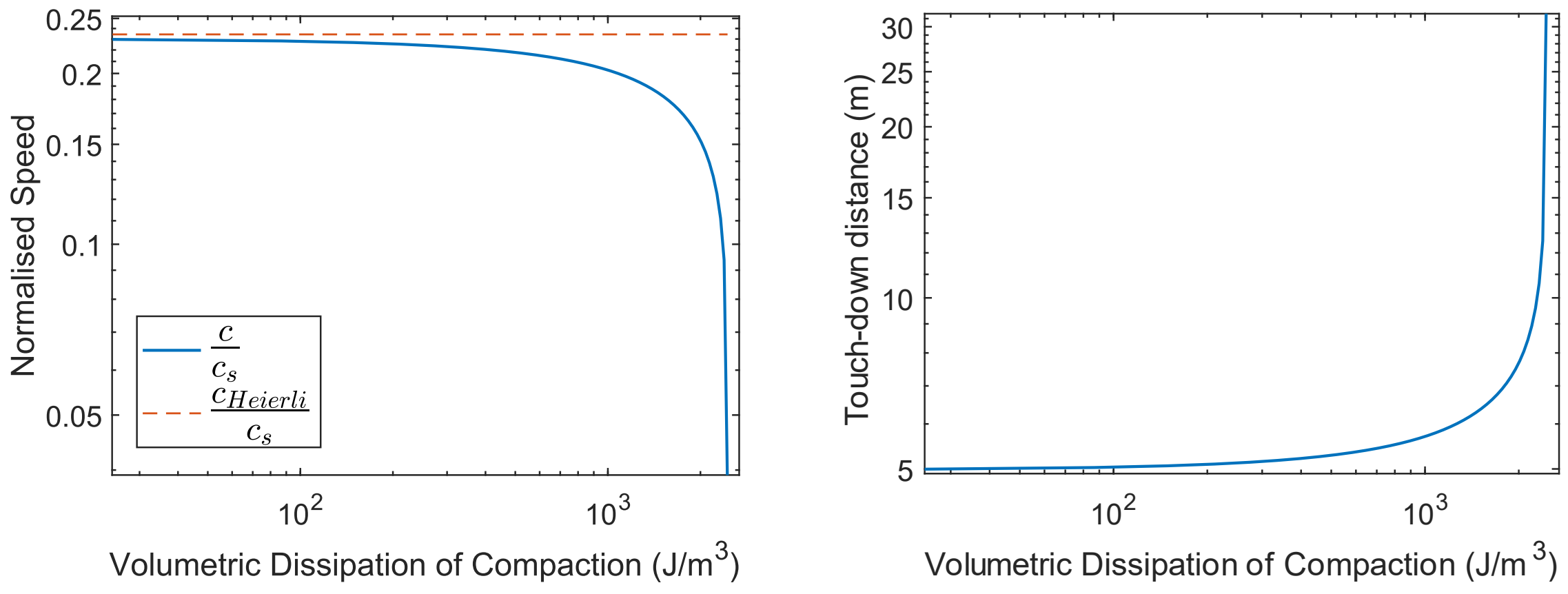

FIGURE 9 – Anticrack speed $c$ (**left**) and touch-down distance $L$ (**right**) as a function of the volumetric dissipation of compaction. *Reading*: $c_{Heierli}$ refers to the speed out from Equation (15), $c_s$ denotes the shear plane wave speed. *Note*: high values of compaction energy are unlikely, yet plotted to highlight the behaviour of the velocity across the entire range of effective acceleration of gravity. *Parameters*: $\rho = 250\ kg.m^{-3},\ \ E = 10\ MPa,\ \ \nu = 0.3,\ \ \ b = h = 1\ m,\ \ g =$ 9,806 $m^2.s^{-1},\ \theta = 0$.

## B. Supershear regime

This section focuses on the *supershear* propagation regime of the crack. It is characterised by crack velocities greater than the slab shear S-waves speed $c_s$, and bounded by the slab compressional P-waves speed $c_p$.

Section VI of Supplementary Material n°2 contains a proof that supershear speeds are inaccessible for transverse bending perturbations. Thus, when observed, the transition from a sub-Rayleigh to a supershear speed range is necessarily accompanied by a change in the mode of propagation, from a transverse to a purely longitudinal perturbation. The Burridge-Andrews mechanism justifies the transition to supershear velocities by the spontaneous nucleation, under the effect of strong tensile gravitational stress in the slab, of another crack upstream from the initial anticrack (Burridge, 1973) (Andrews, 1976). This *daughter crack* then necessarily propagates in pure shear, much faster than the initial anticrack. We thus set up a "strength-of-materials" model for this longitudinal supershear disturbance, which will convey most of the information and energy. The new situation and its parameters are presented in Figure 10.

After the first developments in Supplementary Material n°3, postulating again the acceleration at the crack tip seems unavoidable, leading us to the following strategy: we generalise the boundary condition applied by Heierli (2005) for flexural waves, although it now involves a point force. When it exists, the latter is associated with the energy previously stored in the restoring force of the weak layer and suddenly released when the weak layer fails at the tip: provided that it is instantaneously recovered by the slab, it results in an "acceleration boost" which is expected to help increase the speed of the tip.

Starting from here, the following form for the anticrack speed is found:

$$c = \frac{c_p}{\sqrt{2}}\sqrt{\frac{\left(1 + \frac{\tau_b}{\tau_g - \tau_r}\right)}{\left(1 + \frac{1}{2}\frac{\tau_b}{\tau_g - \tau_r}\right)}} \tag{17}$$

where $\tau_b$ denotes the "boost stress" which can be taken equal to $\tau_p = G_{WL}\frac{u_p}{D_{WL}}$ in case of a brittle fracture (see Supplementary Material n°3 which, in addition, discusses the above assumptions).

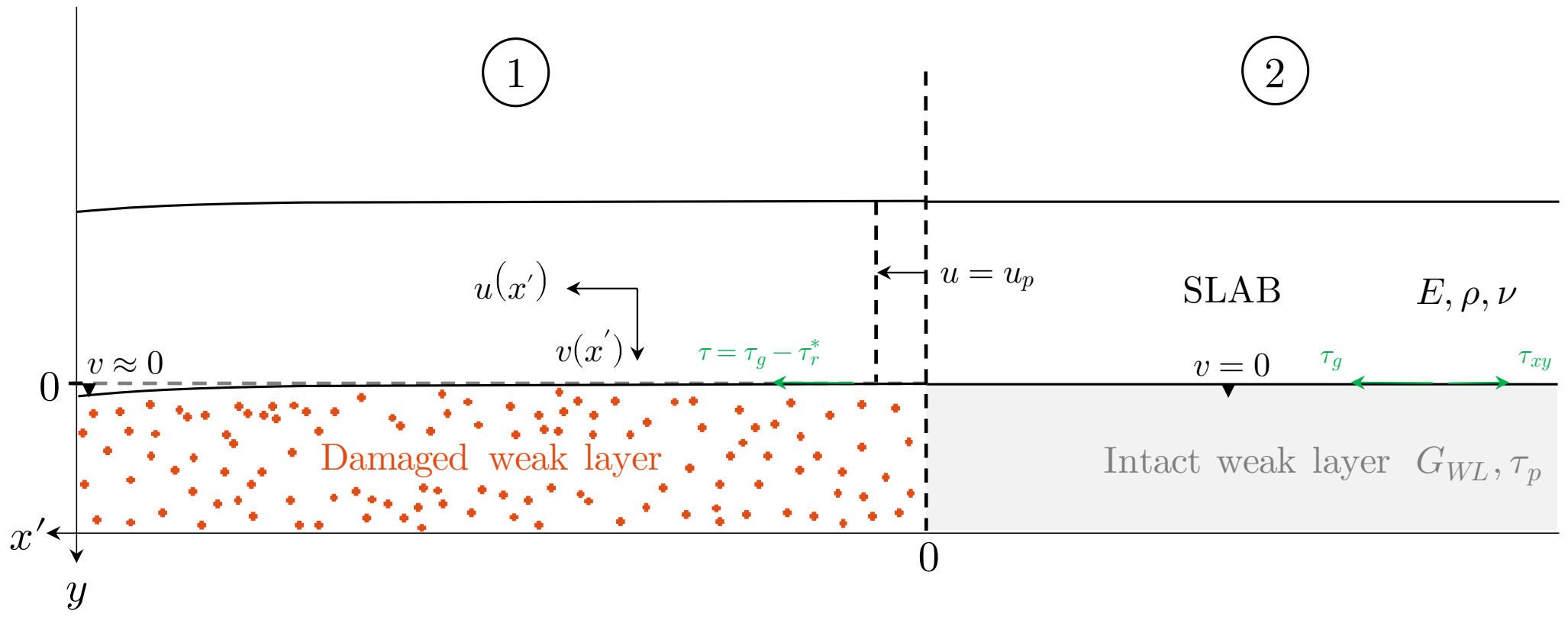

FIGURE 10 – Division of the slab into two sections in the moving reference frame in the *supershear* regime. (*Section 2*) From $-\infty$ to $x' = 0$, the beam is supported by the weak layer stretched elastically in shear under the effect of the slab tension due to gravity. The shear resistance is denoted $\tau_{xy}$. At $x' = 0$, the maximum shear strain $u_p$ is reached and the weak layer abruptly breaks. (*Section 1*) From $x' = 0$ to $+\infty$, the slab is subject, along $x'$, only to gravity $\tau_g$ and to a constant residual friction $\tau_r$ with the debris of the damaged weak layer, lower than gravity. *Note*: the transverse displacement $v$ and the angle $\psi$ are neglected. Besides, the diagram is not to scale.

As expected, the "boost" stress $\tau_b$, which accounts for the restoring energy transferred from the weak layer to the slab at the tip when the fracture happens, does have the effect of increasing the velocity in the proportions defined by the previous ratio. When $\tau_b$ goes to nil, we recover a similar case to the one adopted by Heierli (2005) in the sub-Rayleigh regime, for which gravity (here reduced by friction) is the only force acting at the crack tip, leading to $c = \frac{c_p}{\sqrt{2}}$ which is the lower bound of this formula. Conversely, in the limit where $\tau_b$ grows such that $\frac{\tau_b}{\tau_g - \tau_r}$ becomes predominant over 1, $c \to c_p$, so that we recover an upper bound where the speed does not depend on any property of the two layers. Figure 11 plots Equation (17) as a function of the "boost stress" $\tau_b$.

Note that, in this model, only geometrical properties of the weak layer (namely, its height) affect the velocity: the solution is independent on the geometry of the slab. The slope angle is hidden in the slope-parallel component of the gravity stress $\tau_p$, such that it is expected to play a minor role in the formula. Nonetheless, in the absence of "boost" restoring energies, the speed is:

- To converge towards $c_p/\sqrt{2}$;
- Independent on all parameters of the weak layer, especially on its height, and on those of the slab.

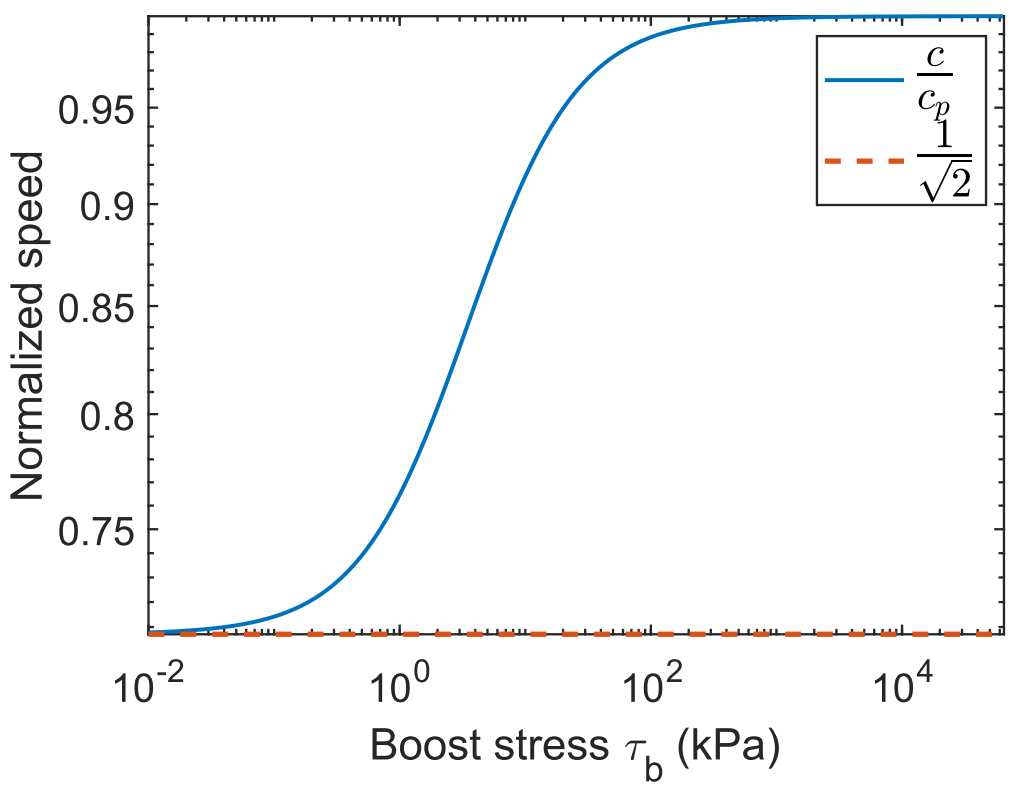

FIGURE 11 – Anticrack velocity as a function of the boost stress $\tau_b$ (*blue curve*), compared to the value $\frac{c_p}{\sqrt{2}}$ found without considering the fracture process of the weak layer at the crack tip (*red curve*). Parameters: $\rho = 250\ kg.m^{-3}$, $E = 3\ MPa$, $\nu = 0.3$, $g = 9.81\ m.s^{-2}$, $h = b = 1m$, $\theta = 0$.

# III - Discussion

## A. Sub-Rayleigh regime

### 1. Speed adequation

The model presented here, which defines itself as an extension of the Heierli (2005) model, has the interest of alleviating the mathematical anomaly that weighed on the original formula, namely the possible divergence of the velocity according to the values given to the physical parameters. Compared to the original *Heierli solution*, the speeds that stem from our generalised model remain bounded between 0 and the limiting S-waves speed $c_s$, regardless of the range of physical constants involved, as shown in Figure 12 for the collapse height of the weak layer.

The generalised solution therefore proves to be a significant improvement of the *Heierli solution*.

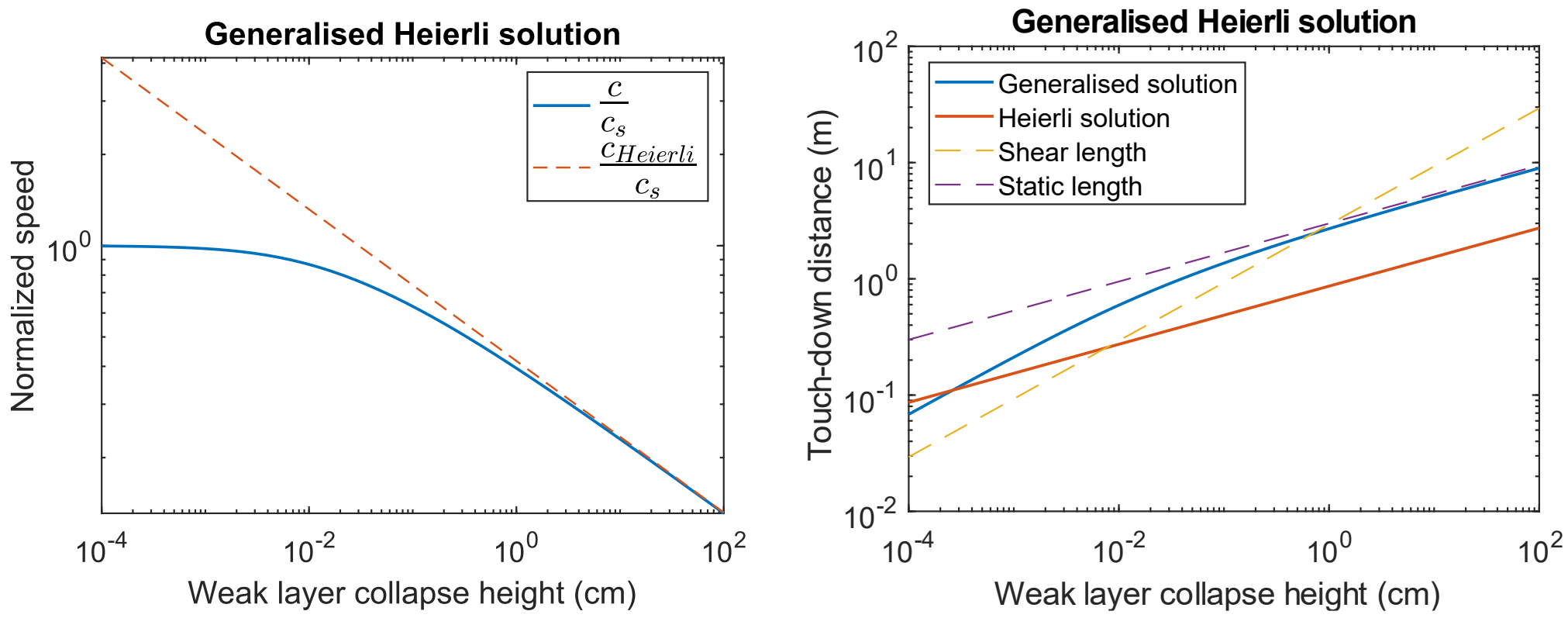

Figure 12 – (**a**) **Left**: Anticrack velocity as a function of the collapse height for our solution (*blue curve*), compared to the Heierli solution (*red curve*), in the absence of fracture energy. (**b**) **Right**: Associated touch-down distance (or *flexural length*). Parameters: $\rho = 250\ kg.m^{-3}$, $E = 3\ MPa$, $\nu = 0.3$, $g = 9.81\ m.s^{-2}$, $h = b = 1m$, $\theta = 0$, $w^{vol}_{comp} = 0$.

### 2. Flexural length mismatch

It can be inferred from Figure 12 that the qualitative behaviour of the *Generalised Heierli solution* is correct as pertains to the touch-down distance: namely, the flexural length increases with the collapse height, since the free fall duration of the slab is maintained constant over the central section, whereas the height until touch-down is increasing.

However, the *Heierli solution*, and our contribution to this approach, still have a fundamental flaw related to the value given to the boundary condition on the curvature at $x' = L$. Indeed, when the speed becomes really low ($c \to 0$) or high ($c \to c_s$), the flexural length is expected to decrease (respectively, to grow) up to a *static length* $L_0$ (respectively, a *shear length* $L_s$) which stems from Equation (8) with a parameterization $c = 0$ ($c = c_s$, respectively). These conditions of junction between boundary speed values, which are stated on the touch-down distance $L$, translate into a condition on the curvature $v^{(2)}(L)$:

$$v^{(2)}(L) \underset{c\to 0}{\longrightarrow} h\frac{L_0}{6}A'_0 \qquad\qquad v^{(2)}(L) \underset{c\to c_s}{\longrightarrow} h\lambda^2 A'_0$$

with $A'_0 = \frac{\Sigma_{eff}}{(h\lambda)^2}$, $L_0 = \sqrt[4]{\frac{72 \cdot h_c \cdot h \cdot \lambda^2}{\Sigma_{eff}}}$. The complete derivation of these terms can be found as Supplementary Material n°5.

The acceleration at the crack tip *cannot be taken as a constant* with respect to the speed, and in peculiar, it cannot be the constant acceleration of gravity $g$ that Heierli (2005) had considered. This justifies the clear mismatch in Figure 12 (right) where the touch-down distance cannot recover its static and shear values for low and high speeds, respectively. Note that, although it is quantitatively wrong, the *Generalised Heierli solution* still shows a correct qualitative behaviour as pertains to its limiting cases, with a dependence growing as $\sqrt[4]{h_c}$ for low speeds and as $\sqrt[2]{h_c}$ for high ones; it is another enhancement to the *Heierli solution*.

Nevertheless, we found no natural form of interpolation between the two boundary values. From this point on, in the absence of knowledge on this constraint, the latter should be avoided, meaning that all velocities would remain accessible by the system. At best, a dispersion curve between a velocity and the corresponding bending length $L$ could be given (Supplementary Material n°7):

$$L^2 = \frac{2hh_c}{\Sigma_{eff}} \frac{c^2}{c_s^2} + r^2(1-\cos(\gamma)) \left(1 - \frac{\gamma \sin(\gamma)}{1-\cos(\gamma)} + \frac{\gamma\cos(\gamma) - \sin(\gamma)}{\gamma - \sin(\gamma)}\right) \tag{18}$$

with $\gamma = \frac{L}{r}$ and $r = \lambda \frac{c}{c_s} \sqrt{\left(1-\frac{c^2}{c_s^2}\right)\left(1-\frac{c^2}{c_p^2}\right)}$. This relation still satisfies the limits:

- For $c \approx c_s$, $L \sim \sqrt{\frac{2hh_c}{\Sigma_{eff}}} \equiv L_s$
- For $c \approx 0$, $L \sim \sqrt[4]{\frac{72hh_c\lambda^2}{\Sigma_{eff}}} \equiv L_0$

## 3. Beyond compaction energy

Our model, which combines fracture mechanics and strength of materials, fails to provide a complete description of the phenomena involved: it requires to choose a specific value of curvature at the fracture tip *ex nihilo* - which, besides, proves to be in our case not admissible at high and low speeds. The curvature should probably not be forced, but rather obtained *via* a dynamical weak layer model, that makes use of the strength of materials to uncover a constitutive law for the still intact weak layer (Section n°3 of Figure 5). Such a model has been put forward by Rosendahl and Weißgraeber (2020a) in a static framework, and we generalised it to a dynamic configuration in Supplementary Material n°4. In short, this model postulates that the weak layer is deformable and undergoes a brittle fracture at the anticrack tip. Although the complexity of the equations would require to solve them numerically, which is not the purpose here, it is still interesting to look at them for two reasons:

- On the one hand, the inclusion of elasticity in this model could explain the oscillations which are sometimes detected in the signals of acceleration in experiments and numerical simulations of Propagation Saw Tests (see, for instance, Figures B.5 and 3.7c in (Bobillier, 2022)).

- On the other hand, the boundary condition $v'(L) = 0$ retained until now appears to be compromised. This hypothesis meant that the slab had to horizontally tangent the weak layer at the anticrack tip. However, the model suggests that the value of the tangent is strictly positive and determined by both weak layer and slab properties. Figure 13 plots the speed obtained by solving equations (9) and (10) for increasing values of the tangent at the crack tip. It shows that higher speeds can be recovered from this reasoning, growing from that of the *Heierli solution* up to the limit shear speed $c_s$. This seems more in line with the simulations when considering no fracture nor compaction (Trottet, Simenhois, Bobillier, van Herwijnen, et al., 2022).

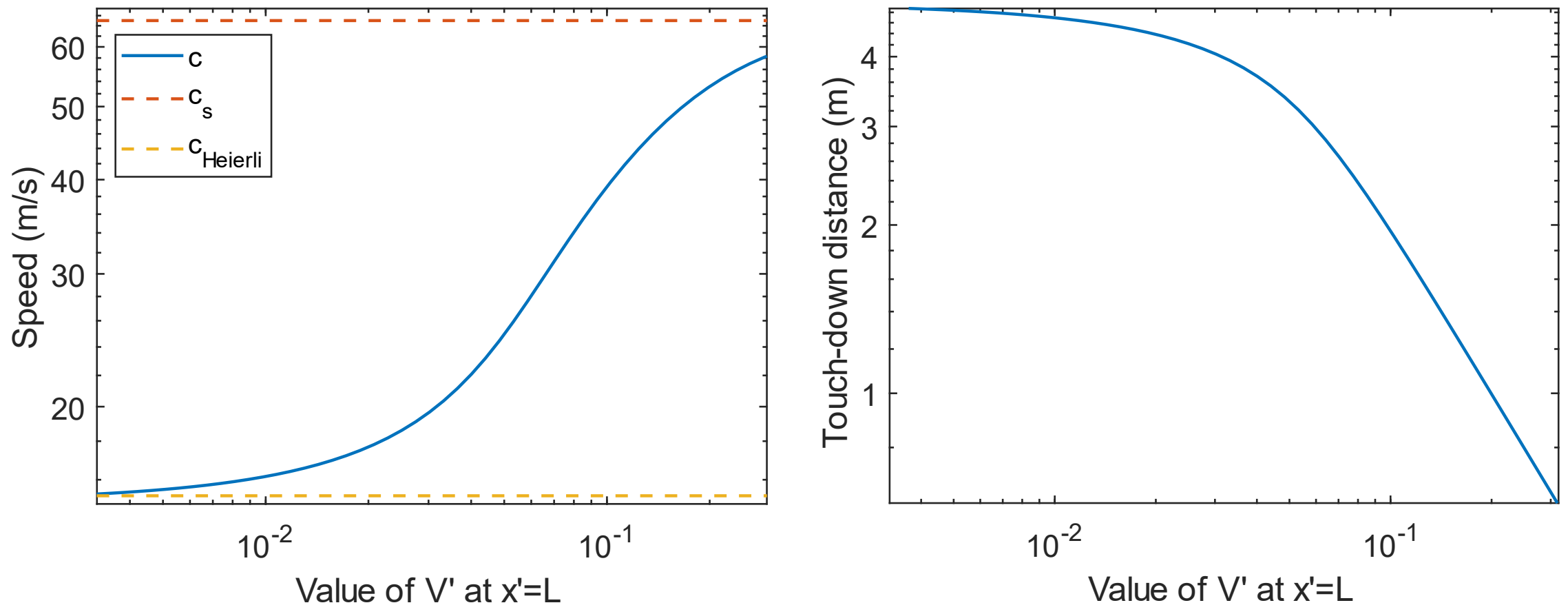

FIGURE 13 – (**a**) **Left**: Anticrack velocity as a function of the tangent at the crack tip (*blue curve*), compared to the Heierli solution (*yellow*) and $c_s$ (*red*), in the absence of fracture energy. (**b**) **Right**: Associated touch-down distance. Parameters: $\rho = 250\ kg.m^{-3}$, $E = 3\ MPa$, $\nu = 0.3$, $g = 9.81\ m.s^{-2}$, $h = b = 1m$, $\theta = 0$, $w_{comp}^{vol} = 0$, $h_c = 1\ cm$.

## B. Supershear regime

It is not obvious at first sight whether Equation (17) is valid or not, since there are very few experimental data available as for now on the supershear transition in slab avalanches. The main numerical data set on which we rely is that of Trottet et al. (2022). Once again, let $c$ denote the speed of the crack in its steady state. Trottet et al. (2022) observed that $c \approx\ 1.6 \cdot c_s \sim c_p$, where the latter order of magnitude is obtained by recalling that $\frac{c_p}{c_s} = \sqrt{\frac{E}{G}} = \sqrt{\frac{2(1-\nu)}{1-2\nu}} \approx 1.9$, using a typical value of 0.3 for $\nu$. Without considering the fracture process at the crack tip, our formula simplifies to $c = \frac{c_p}{\sqrt{2}} \sim 1.2 \cdot c_s$. By noticing that our P-wave speed $c_p$ is defined here for a *two-dimensional* model, and is therefore *not equal* to the three-dimensional dilatational speed appearing in Trottet et al. (2022), a careful comparison would nonetheless lead us to the conclusion that the observed speeds are either of the order of, or greater, than the lower limit we found (namely, $c = \frac{c_p}{\sqrt{2}}$).

Based on our modelling, we note that the boundary condition on the curvature $u^{(2)}$ at the anticrack tip was taken constant and equal to gravity; this hypothesis has been proven wrong in the sub-Rayleigh regime and is therefore likely incorrect, thereby probably leading to an incorrect speed indeed. By considering "rudimentarily" the fracture process at the crack tip in the form of a point force contributing to the curvature, Equation (17) showed that higher speeds could be recovered. In our reasoning, however, it does not seem possible to thoroughly derive the curvature at the anticrack tip from what happens on the still-intact portion of the weak layer, even through an utter strength-of-materials perspective (see equations in Supplementary Material n°4). Consequently, either a more complex modelling of the weak layer or another analytical approach could be needed for further investigations.

# Conclusion & Outlook

This paper developed a new model for dynamic anticrack propagation based on the Timoshenko beam kinematics. Our speed formulas for the steady-state anticrack, available for both the *sub-Rayleigh* and the (newly discovered) *supershear* regime, bypass the main problems of the previously available one provided by Heierli (2005). In particular, the predicted speeds remain bounded for all values of the physical parameters involved, and consider the compaction of the weak layer derived from the perspective of fracture mechanics. For typical values of snow parameters in the slab and the weak layer, the resulting speeds show qualitatively good agreement with the available data. Quantitatively, however, they demonstrate fundamental flaws linked to the chosen boundary conditions, especially with respect to the touch-down distance for anticracks. Besides, the fracture processes at the anticrack tip were not accounted for. A more accurate model for the failure upstream from the anticrack could help explain the limitations of the present study and push the knowledge of the process a step further. Although incomplete due to the inherent mathematical difficulties encountered while deriving slightly more complex models - especially when attempting to regard fractures at the anticrack tip -, this paper underlines the robustness of our computation strategy, which could lead to an improved understanding of anticracks if more experimental data could be accessed.

The implications of this study are at least twofold: greater knowledge of the speed of an anticrack could lead not only to greater accuracy in estimating the size of avalanches when released, but also to a better estimate of the risk of their triggering. This last aspect is emphasised by Bergfeld et al. (2022) when proposing a new stability index, the *SSP* (Index for Self-Sustained Propagation), based on both static (the critical crack length) and dynamic indicators (such as the anticrack speed).

Our results do not reflect the transient dynamics of the anticrack; they especially give no information with respect to the convergence towards the steady state as a function of the weak layer parameters, on which the *supershear* regime seems to drastically depend (Trottet et al., 2022). Based on this observation, the transient dynamics of the collapse could be sought from a generalization of the expressions established by Heierli (2008) from a static perspective of the problem, whether this is achieved by (i) the differentiation of the action functional resulting from the addition of a kinetic term to the static potential of the anticrack found by Heierli (2008) in equation (4.10), or (ii) by directly searching for a dynamic potential for the anticrack following the approach that led, in the static case, to equation (4.10).

Please finally note that we have chosen to derive mixed models (involving both fracture mechanics and the strength of materials) or strength-of-materials models only. Adopting a framework utterly based on fracture mechanics could also prove to be a valuable alternative, by adapting to snow the extensive literature on transient dynamic fracture mechanics. To this end, one could draw inspiration from the works of Marder (1991) Freund (1998) and Svetlizky et al. (2017) for *anticracks*, and of Kammer et al. (2018), Bayart et al. (2018) and Shlomai et al. (2020) for the *supershear* cracks.

# Acknowledgments

We acknowledge Grégoire Bobillier, Bastian Bergfeld and Alec van Herwijnen for helpful discussions on anticrack propagation, and for having provided us with numerical and experimental data which refined our investigations. We are grateful to Elsa Bayart for valuable discussions on the approach of fracture mechanics in both *sub-Rayleigh* and *supershear* regimes. This research has been supported by the Swiss National Science Foundation (grant n° PCEFP2_181227).

*Author contributions*: The authors contributed equally to the model, from the conception of the methodology up to the leading physical assumptions of the analytical part. M.S. performed the derivations and drafted the paper under the supervision of B.T. and J.G. J.G. defined the framework of the study and obtained the funding. All authors contributed equally to the final version of the paper.

*Competing interests*: The authors declare no competing interests.

*Supplementary Information*: Supplementary Materials (numbered 1 to 8) are enclosed with this paper.

# Supplementary materials:

# A theoretical framework for dynamic Anticrack and Supershear propagation in snow slab avalanches

Marin SIRON[1,2], Bertil TROTTET[1], Johan GAUME[3,4,5]

[1] School of Architecture, Civil and Environmental Engineering, Swiss Federal Institute of Technology, Lausanne, Switzerland
[2] Ecole Normale Supérieure de Lyon, Université Claude Bernard, Lyon, France
[3] Institute for Geotechnical Engineering, ETH Zurich, Zurich, Switzerland
[4] WSL Institute for Snow and Avalanche Research SLF, Davos Dorf, Switzerland
[5] Climate Change, Extremes, and Natural Hazards in Alpine Regions Research Center CERC, Davos Dorf, Switzerland

***Content***

Statement of Hamilton's principle

The motion of a conservative elastic solid between two instants $t_1$ et $t_2$ takes place in such a way that the functional $S(\{u_i\}, t) = \int_{t_1}^{t_2} \mathcal{L}(\{u_i\}, t)\, dt$ is stationary:

$$\delta S = \delta \int_{t_1}^{t_2} \mathcal{L}(u_1, \dots, u_n, t)\, dt = 0 \qquad\qquad \forall \delta u_i \neq 0$$

and this, for any non-zero virtual variation $\delta u_i$ ($i \in [0, n]$) of the $n$ fields of displacement satisfying:

- The kinematic boundary conditions: $u_i(\vec{r}, t) = u_i^{\Sigma}(\vec{r}, t) \Leftrightarrow \delta u_i(\vec{r}, t) = 0$ for all $\vec{r} \in \Sigma$ (where $u_i^{\Sigma}(\vec{r}, t)$ denote the displacements imposed on the boundary $\Sigma$);
- The initial conditions $\delta u_i = 0$ for $t = t_1$ and $t = t_2$.

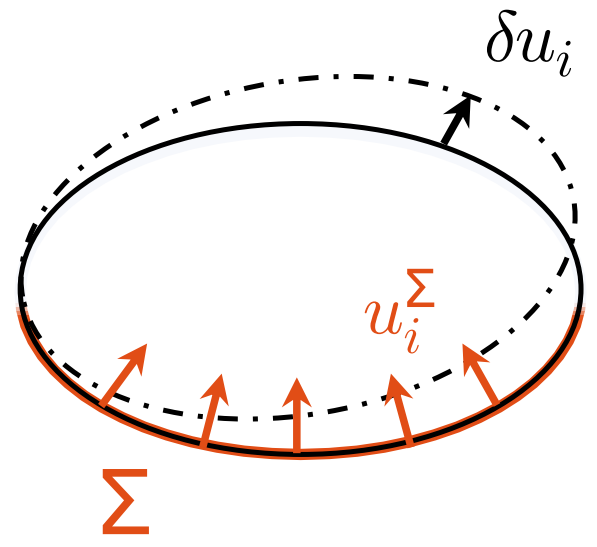

The boundary conditions in constraints are imposed by Hamilton's principle to ensure the stationary action condition.

Using Hamilton's principle to derive the equations of motion

This approach shows two advantages here:

- An energetic approach is appropriate for this problem, since the state of the art has so far formulated the volumetric collapse of the weak layer in terms of *fracture energy* required to fracture the snow. Furthermore, the energy quantities can be easily integrated over the whole slab in a scalar approach. The use of analytical mechanics (Euler-Lagrange equations or variational principle) is therefore advisable.

- Compared to the Euler-Lagrange equations directly using the Lagrangian, Hamilton's principle manipulates its time integral $\mathcal{S}$ (the action) between times $t_1$ and $t_2$, which it then minimises along any virtual path $\delta(u, v, \psi)$. In doing so, once the kinematic boundary conditions and the initial conditions (known to the user) are defined, the action extremum conditions impose the stress boundary conditions, the knowledge of which is necessary to close the system of equations. This is a major difference with respect to the Euler-Lagrange equations: Hamilton's principle provides the user with the stress conditions at the boundaries, which have to be postulated in the opposite case.

Please note, however, that the system ought to be conservative for this formulation of the variational calculus to be valid. In practice, our system is not conservative since the propagation of the subsidence is irreversible, mainly due to frictional forces between the slab and the weak layer and to plastic strains during the weak layer compaction. Yet, since only steady-state wave propagation is as stake, it is possible to reduce to a conservative system and use the solutions of this approach.

# Supplementary Material n°2 - Deriving the equations of motion in the Sub-Rayleigh regime

In the rest of the document, $\cdot$ denotes time derivatives and $'$ denotes spatial derivatives.

## I. Dynamic equations for the motion of the disturbance

To begin with, let us place ourselves in the fixed observational reference frame between two arbitrary instants 0 and $t$, on a section $[x_1, x_2]$ of the disturbance, such that the beam is bending between these two instants. The action functional is expressed as a function of all external forces at $x_1$ and $x_2$ respectively (namely the normal force $N_{x_1,x_2}$, shear force $Q_{x_1,x_2}$ and bending moment $\mathcal{M}_{x_1,x_2}$):

$$S(u,v,\psi,t) = b\cdot h\int_0^t \left\{ \int_{x_1}^{x_2} \left\{ \frac{1}{2}\rho\dot{u}^2 + \frac{1}{2}\rho\dot{v}^2 + \frac{1}{2}\rho I\dot{\psi}^2 - \frac{1}{2}Eu'^2 - \frac{1}{2}EI\psi'^2 - \frac{1}{2}\kappa G(v'-\psi)^2 + \frac{\tau}{h}u - \frac{\sigma}{h} - \frac{w_{comp}^{vol}}{h}v \right\} dx - \frac{1}{bh}\left(N_{x_1,x_2}v + Q_{x_1,x_2}u + \mathcal{M}_{x_1,x_2}\psi\right) \right\} dt \tag{1}$$

$$\Downarrow$$

$$\frac{\delta S}{bh} = \int_0^t \left\{ \int_{x_1}^{x_2} \left\{ \rho\dot{u}\cdot\delta\dot{u} + \rho\dot{v}\cdot\delta\dot{v} + \rho I\dot{\psi}\cdot\delta\dot{\psi} - Eu'\delta u' \cdot - EI\psi'\delta\psi' - \kappa G(v'-\psi)\delta(v'-\psi) + \frac{\tau}{h}\delta u - \frac{1}{h}(\sigma + w_{comp}^{vol})\delta v \right\} dx - \frac{1}{bh}\left(N_{x_1,x_2}\delta v + Q_{x_1,x_2}\delta u + \mathcal{M}_{x_1,x_2}\delta\psi\right) \right\} dt \tag{2}$$

Integrating by parts over $t$ and $x$, respectively, and defining $\sigma_{eff} = \sigma + w_{comp}^{vol}$, we get:

$$\begin{aligned} \frac{\delta S}{bh} = & \int_0^t \left\{ \int_{x_1}^{x_2} \left\{ -\rho\ddot{u}\cdot\delta u - \rho\ddot{v}\cdot\delta v - \rho I\ddot{\psi}\cdot\delta\psi + Eu''\delta u + EI\psi''\delta\psi + \kappa G(v'-\psi)\delta\psi + \kappa G(v''-\psi')\delta v + \frac{\tau}{h}\delta u - \frac{\sigma_{eff}}{h}\delta v \right\} dx \right\} dt \\ & + \int_{x_1}^{x_2} \{ [\rho\dot{u}\delta u]_0^t + [\rho\dot{v}\delta v]_0^t + [\rho I\dot{\psi}\delta\psi]_0^t \} dx \\ & - \int_0^t \{ [(Eu' - Q_{x_1,x_2})\delta u]_{x_1}^{x_2} + [(EI\psi' - \mathcal{M}_{x_1,x_2})\delta\psi]_{x_1}^{x_2} + [(\kappa G(v'-\psi) - Q_{x_1,x_2})\delta v]_{x_1}^{x_2} \} dt \end{aligned} \tag{3}$$

$$\begin{aligned} \frac{\delta S}{h} = & \int_0^t \int_{x_1}^{x_2} \left\{ \left(-\rho\ddot{u} + Eu'' + \frac{\tau}{h}\right)\delta u + \left(-\rho\ddot{v} + \kappa G(v''-\psi') - \frac{\sigma_{eff}}{h}\right)\delta v + \left(-\rho I\ddot{\psi} + +EI\psi'' + \kappa G(v'-\psi)\right)\delta\psi \right\} dx\, dt \\ & + \int_{x_1}^{x_2} \{ [\rho\dot{u}\delta u]_0^t + [\rho\dot{v}\delta v]_0^t + [\rho I\dot{\psi}\delta\psi]_0^t \} dx \\ & - \int_0^t \left\{ \left[\left(Eu' - \frac{Q_{x_1,x_2}}{bh}\right)\delta u\right]_{x_1}^{x_2} + \left[\left(EI\psi' - \frac{\mathcal{M}_{x_1,x_2}}{bh}\right)\delta\psi\right]_{x_1}^{x_2} + \left[\left(\kappa G(v'-\psi) - \frac{Q_{x_1,x_2}}{bh}\right)\delta v\right]_{x_1}^{x_2} \right\} dt \end{aligned} \tag{4}$$

Using Hamilton's principle gives:

- For all virtual displacement fields which are kinetically admissible, $\delta u = \delta v = \delta\psi = 0$ at both moments 0 and $t$.
  => $[\rho\dot{u}\delta u]_0^t = [\rho\dot{v}\delta v]_0^t = [\rho I\dot{\psi}\delta\psi]_0^t = 0$
- The condition of stationary action functional $\delta S = 0$ provides the following differential equations:

$$\begin{cases} Eu'' - \rho\ddot{u} = -\dfrac{\tau}{h} \\ \kappa Gv'' - \rho\ddot{v} - \kappa G\psi' = \dfrac{\sigma_{eff}}{h} \\ EI\psi'' - \rho I\ddot{\psi} - \kappa G\psi + \kappa Gv' = 0 \end{cases} \tag{5}$$

– The stress boundary conditions read:
$\forall t,$

$$\int_0^t \left[\left(Eu' - \frac{Q_{x_1,x_2}}{bh}\right)\delta u\right]_{x_1}^{x_2} dt = \int_0^t \left[\left(EI\psi' - \frac{\mathcal{M}_{x_1,x_2}}{bh}\right)\delta\psi\right]_{x_1}^{x_2} dt = \int_0^t \left[\left(\kappa G(v'-\psi) - \frac{Q_{x_1,x_2}}{bh}\right)\delta v\right]_{x_1}^{x_2} dt = 0$$
$$\Downarrow$$

$$\int_0^t \left(Eu'(x_2,t) - \frac{N_{x_2}}{bh}\right)\delta u(x_2,t)dt = \int_0^t \left(Eu'(x_1,t) - \frac{N_{x_1}}{bh}\right)\delta u(x_1,t)dt = 0$$
$$\int_0^t \left(EI\psi'(x_2,t) - \frac{\mathcal{M}_{x_2}}{bh}\right)\delta\psi(x_2,t)dt = \int_0^t \left(EI\psi'(x_1,t) - \frac{\mathcal{M}_{x_1}}{bh}\right)\delta\psi(x_1,t)dt = 0$$
$$\int_0^t \left(\kappa G(v'-\psi)(x_2,t) - \frac{Q_{x_2}}{bh}\right)\delta v(x_2,t)dt = \int_0^t \left(\kappa G(v'-\psi)(x_1,t) - \frac{Q_{x_1}}{bh}\right)\delta v(x_1,t)dt = 0$$

Thus, the kinematic boundary conditions determine the stress boundary conditions. For example, let us assume the complete absence of kinematic boundary conditions. Then the kinematically admissible displacements must verify $\delta u, \delta v, \delta\psi \neq 0$ over time at the positions $x_1$ and $x_2$ of the edges. The six stress equations are then determined by:

$$\begin{aligned}
Eu'(x_1,t) &= \frac{N_{x_1}}{bh}\\
Eu'(x_2,t) &= \frac{N_{x_2}}{bh}\\
\kappa G(v'-\psi)(x_2,t) &= \frac{Q_{x_2}}{bh}\\
\kappa G(v'-\psi)(x_1,t) &= \frac{Q_{x_1}}{bh}\\
EI\psi'(x_1,t) &= \frac{\mathcal{M}_{x_1}}{bh}\\
EI\psi'(x_2,t) &= \frac{\mathcal{M}_{x_2}}{bh}
\end{aligned} \tag{6}$$

Obviously, if $n \leq 6$ kinematic boundary conditions are set on $u, v$ or $\psi$ in $x_1$ and/or $x_2$, this erases the associated $n$ stress boundary conditions, since the kinematic boundary conditions require $\delta u, \delta v$ or $\delta\psi$ to be zero in $x_1$ (respectively in $x_2$).

Later on, it will be more convenient to work on dimensionless fields. We define:

$U = \frac{u}{h}$ ; $V = \frac{v}{h}$ ; $\Sigma_{\text{eff}} = -\frac{\sigma_{eff}}{\kappa \mathrm{G}}$ ; $\mathrm{T} = \frac{\tau}{\kappa G}$ ; $\lambda = \sqrt{\frac{EI}{\kappa G}}$ ; $c_s = \sqrt{\frac{\kappa G}{\rho}}$ ; $c_p = \sqrt{\frac{E}{\rho}}$

so that the dimensionless system reads:

$$\begin{aligned}
&U'' - \frac{1}{c_p^2}\ddot{U} = -\frac{\kappa G}{Eh^2}T = -\frac{T}{12}\frac{1}{\lambda^2}\\
&V'' - \frac{1}{c_s^2}\ddot{V} - \frac{1}{h}\psi' = -\frac{\Sigma_{\text{eff}}}{h^2}\\
&\psi'' - \frac{1}{c_p^2}\ddot{\psi} - \frac{1}{\lambda^2}\psi + \frac{h}{\lambda^2}V' = 0
\end{aligned} \tag{7}$$

## II. Stationary equations of motion for the disturbance

Now we look for *propagative stationary* solutions of these equations moving towards *increasing 'x'*. They are of the well-known form: $U(x,t) = U(x-ct),\ V(x,t) = V(x-ct),\ \psi(x,t) = \psi(x-ct)$. For this reason, the system of equations can be re-written entirely using the new variable $x' = x - ct$, for which:

$$\partial x = \partial x'$$
$$\partial t = -\frac{1}{c}\partial x'$$

By noting $\cdot^{(k)} = \frac{\partial^k}{\partial x'^k}$, the system reads:

$$\left(1 - \frac{c^2}{c_p^2}\right) U^{(2)} = -\frac{\kappa G}{Eh^2} T \tag{8}$$
$$\left(1 - \frac{c^2}{c_s^2}\right) V^{(2)} - \frac{1}{h}\psi^{(1)} = -\frac{\Sigma_{\text{eff}}}{h^2} \tag{9}$$
$$\left(1 - \frac{c^2}{c_p^2}\right) \psi^{(2)} - \frac{1}{\lambda^2}\psi + \frac{h}{\lambda^2} V^{(1)} \tag{10}$$

These equations are valid only *in the moving reference frame* attached to the perturbation, on the strictly disturbed section of the beam (referred to as $]0, L[$). Two conventions are chosen regarding the edges:

- The break occurs at $x' = L$: it is instantaneous and involves a reaction $\vec{f} = \overrightarrow{f_I} + \overrightarrow{f_{II}}$ of the supporting weak layer, due to fracturing in both shear and compression. Thus, at $x' = L$, the beam is subject to gravity in the same way as the rest of the section $]0, L[$, but also to the fracture reaction of the weak layer.
- The $x' = 0$ point is defined as the point of tangency, i.e., the point at and below which the beam rests on the substrate. In other words, at $x' = 0$, the beam is not subject to gravity, because the latter is compensated by the reaction of the substrate.

To allow for the junction between the unsupported and supported sections of the slab (at $x' = 0$ and $x' = L$), the displacements must be continuous, translating to:

$$u(0^-,t) = u(0^+,t) \qquad v(0^-,t) = v(0^+,t) \qquad \psi(0^-,t) = \psi(0^+,t)$$
$$u(L^-,t) = u(L^+,t) \qquad v(L^-,t) = v(L^+,t) \qquad \psi(L^-,t) = \psi(L^+,t)$$

The stress boundary conditions in steady state are unchanged compared to the transient regime (eq. (6)) as it is reminded that $\partial x = \partial x'$ (see Supplementary Material n°8); thus, they read:

$$EAu'(0,t) = \mathcal{N}_0 \qquad EIA\psi'(0,t) = \mathcal{M}_0 \qquad \kappa GAv'(0,t) = Q_0$$
$$EAu'(L,t) = \mathcal{N}_L \qquad EIA\psi'(L,t) = \mathcal{M}_L \qquad \kappa GAv'(L,t) = Q_L$$

In effect, the latter conditions express the continuity of the internal forces and moments which are applied to the cross sections of the beams. Since their expression is known on both sides of the singularities, this eventually leads to the same stress boundary conditions as eq.(18) and (19) in Rosendahl & Weißgraeber (2020a):

$$EAu'(0^-,t) = EAu'(0^+,t) \qquad \kappa GAv'(0^-,t) = \kappa GAv'(0^+,t) \qquad -EIA\psi'(0^-,t) = -EIA\psi'(0^+,t)$$
$$EAu'(L^-,t) = EAu'(L^+,t) + f_{II} \qquad \kappa GAv'(L^-,t) = \kappa GAv'(L^+,t) + f_I \qquad -EIA\psi'(L^-,t) = -EIA\psi'(L^+,t) + f_{II}\frac{h}{2}$$

In the steady state, the decoupling of the equations on $V$ and $\psi$ in easier:

(9) is rewritten as:

$$h\left(1-\frac{c^2}{c_s^2}\right)V^{(2)}+\frac{\Sigma_{\text{eff}}}{h}=\boxed{\psi^{(1)}} \tag{11}$$

$\frac{d^2}{dx'^2}$(9) reads:

$$h\left(1-\frac{c^2}{c_s^2}\right)V^{(4)}=\boxed{\psi^{(3)}} \tag{12}$$

$\frac{d}{dx'}$(10) reads:

$$\left(1-\frac{c^2}{c_p^2}\right)\boxed{\psi^{(3)}}-\frac{1}{\lambda^2}\psi^{(1)}+\frac{h}{\lambda^2}V^{(2)}=0 \tag{13}$$

Substituting (12) in (13),

$$\left(1-\frac{c^2}{c_s^2}\right)\left(1-\frac{c^2}{c_p^2}\right)V^{(4)}-\frac{1}{\lambda^2 h}\left[\boxed{\psi^{(1)}}\right]+\frac{1}{\lambda^2}V^{(2)}=0 \tag{14}$$

Substituting (12) in (15),

$$\left(1-\frac{c^2}{c_s^2}\right)\left(1-\frac{c^2}{c_p^2}\right)V^{(4)}-\frac{1}{\lambda^2 h}\left[h\left(1-\frac{c^2}{c_s^2}\right)V^{(2)}+\frac{1}{h}\Sigma_{eff}\right]+\frac{1}{\lambda^2}V^{(2)}=0 \tag{15}$$

which eventually results in Equation (16).

$$\left(1-\frac{c^2}{c_s^2}\right)\left(1-\frac{c^2}{c_p^2}\right)V^{(4)}+\frac{1}{\lambda^2}\frac{c^2}{c_s^2}V^{(2)}=\frac{1}{(\lambda h)^2}\Sigma_{eff} \tag{16}$$

*General solution of $U$ ($c<c_p$):*

$$U(x')=-\frac{1}{2}\frac{T}{12\lambda^2}\frac{1}{1-\frac{c^2}{c_p^2}}x'^2+R_1x'+R_2 \tag{17}$$

*Particular solution of $V$:*

$$V_0^{(2)}(x')=\frac{1}{h^2}\frac{c_s^2}{c^2}\Sigma_{eff}$$
$$V_0(x')=\frac{1}{2h^2}\frac{c_s^2}{c^2}\Sigma_{eff}x'^2+\mathcal{C}_1{}'x'+D_1{}'$$

*General solution of $V$:*

By looking for solutions of the form $e^{kx}$ (where $k$ is a constant), the characteristic polynomial of the differential equation is

$$\left(1-\frac{c^2}{c_s^2}\right)\left(1-\frac{c^2}{c_p^2}\right)k^4+\frac{1}{\lambda^2}\frac{c^2}{c_s^2}k^2=0 \tag{18}$$

Several regimes can be distinguished, depending on the values of $c$, with singularities in $c=c_s$ and $c=c_p$. Outside these singularities, where the particular solution $V_0(x')$ applies as the general solution, 0 is a double root and the other two roots are expressed according to:

$$k^2=-\frac{1}{\lambda^2}\frac{c^2/c_s^2}{\left(1-\frac{c^2}{c_s^2}\right)\left(1-\frac{c^2}{c_p^2}\right)} \tag{19}$$

- If $c\in]c_s,c_p[$ (*supershear* regime), $k^2>0$.

$$k_{\pm} = \pm\frac{c}{\lambda c_s}\frac{1}{\sqrt{\left|\left(1-\frac{c^2}{c_s^2}\right)\left(1-\frac{c^2}{c_p^2}\right)\right|}} = \pm k \equiv \pm\frac{1}{r}$$

The solutions thus turn out to be of real exponential type:

$$V(x') = A_1 e^{\frac{x'}{r}} + B_2 e^{-\frac{x'}{r}} + \frac{1}{2h^2}\frac{c_s^2}{c^2}\Sigma_{eff}x'^2 + \mathcal{C}_1 x' + D_1 \tag{20}$$

- If $c < c_s$ (*sub-Rayleigh* regime), $r^2 < 0$.

$$r_{\pm} = \pm i\frac{c\ c_p}{\lambda}\frac{1}{\sqrt{\left|(c_s^2-c^2)(c_p^2-c^2)\right|}} = \pm i\cdot k = \pm i\frac{1}{r}$$

The solutions thus turn out to be of harmonic type:

$$V(x') = A_2\cos\left(\frac{x'}{r}\right) + B_2\sin\left(\frac{x'}{r}\right) + \frac{1}{2h^2}\frac{c_s^2}{c^2}\Sigma_{eff}x'^2 + \mathcal{C}_2 x' + D_2 \tag{21}$$

*Note:* $r$ defines a characteristic wavelength of the bending disturbance, which decreases non-linearly with the propagation speed $c$:

$$r(c) = \lambda\frac{c_s}{c}\cdot\sqrt{\left|\left(1-\frac{c^2}{c_s^2}\right)\left(1-\frac{c^2}{c_p^2}\right)\right|}\xrightarrow[c\to c_s]{}0$$

*General solution of* $\psi$ :

Using equations $\frac{d}{dx'}(2)$ and (3), and noting that r(c) is defined by the relation $\frac{c^2}{c_s^2}r^2 = \lambda^2\left(1-\frac{c^2}{c_s^2}\right)\left(1-\frac{c^2}{c_p^2}\right)$, one shows that:

$$\psi(x') = h\cdot V^{(1)} + h\lambda^2\left(1-\tfrac{c^2}{c_s^2}\right)\left(1-\tfrac{c^2}{c_p^2}\right)V^{(3)} = h\cdot V^{(1)} + h\tfrac{c^2}{c_s^2}r^2\cdot V^{(3)}$$

Thus,

- If $c \in ]c_s, c_p[$ (*supershear* regime),

$$\psi(x') = \tfrac{h}{r}\left(1-\tfrac{c^2}{c_s^2}\right)\left(A_1 e^{\frac{x'}{r}} - B_1 e^{-\frac{x'}{r}}\right) + \tfrac{1}{h}\tfrac{c_s^2}{c^2}\Sigma_{eff}x' + C_1 h$$
$$\psi^{(1)}(x') = -\tfrac{h}{r^2}\left(1-\tfrac{c^2}{c_s^2}\right)\left(A_1 e^{\frac{x'}{r}} + B_1 e^{-\frac{x'}{r}}\right) + \tfrac{1}{h}\tfrac{c_s^2}{c^2}\Sigma_{eff}$$

- If $c < c_s$ (*sub-Rayleigh* regime),

$$\psi(x') = \tfrac{h}{r}\left(1-\tfrac{c^2}{c_s^2}\right)\left(-A_2\sin\left(\tfrac{x'}{r}\right) + B_2\cos\left(\tfrac{x'}{r}\right)\right) + \tfrac{1}{h}\tfrac{c_s^2}{c^2}\Sigma_{eff}x' + C_2 h$$
$$\psi^{(1)}(x') = -\tfrac{h}{r^2}\left(1-\tfrac{c^2}{c_s^2}\right)\left(A_2\cos\left(\tfrac{x'}{r}\right) + B_2\sin\left(\tfrac{x'}{r}\right)\right) + \tfrac{1}{h}\tfrac{c_s^2}{c^2}\Sigma_{eff}$$

Hence the coupling relations between the displacement fields:

$$\psi(x') = h\left(1-\frac{c^2}{c_s^2}\right)V' + \frac{\Sigma_{eff}}{h}x' + h\frac{c^2}{c_s^2}C_2$$
$$\psi^{(1)}(x') = h\left(1-\frac{c^2}{c_s^2}\right)V'' + \frac{\Sigma_{eff}}{h}$$

Enforcing $\psi$ (respectively, $\psi^{(1)}$) at the edges translates into boundary conditions on $V^{(1)}$ (respectively, $V^{(2)}$). Thus, we expect the boundary conditions to be expressed on $V$, $V^{(1)}$ (or $\psi$) and $V^{(2)}$ (or $\psi^{(1)}$).

## III. Equations of motion of the intact section

The constitutive relation of the Timoshenko beam kinematics in $y = \frac{h}{2}$ reads:

$$u\left(x, \frac{h}{2}, t\right) = u(x,t) - \frac{h}{2}\psi(x,t)$$

On the section which is not yet affected by the disturbance,

- We assume that the slab adheres to the undeformed weak layer at $y = h/2$, imposing $u(x, \frac{h}{2}, t) = 0$. Thus, the condition $u(x,t) = \frac{h}{2}\psi(x,t)$ applies to the whole section.
- The slab rests uniformly on the still intact weak layer. The vertical displacement $v$ is therefore assumed to be zero.

The action functional on an intact portion $[x_3, x_4]$ upstream of the disturbance is subsequently given by:

$$S(\psi, t) = h \int_0^t \int_{x_3}^{x_4} \left\{ \frac{1}{2}\rho \left(\frac{h}{2}\frac{\partial \psi}{\partial t}\right)^2 + \frac{1}{2}\rho I \left(\frac{\partial \psi}{\partial t}\right)^2 - \frac{1}{2}E\left(\frac{h}{2}\frac{\partial \psi}{\partial x}\right)^2 - \frac{1}{2}\kappa G \psi^2 + \frac{\tau}{h}\left(\frac{h}{2}\psi\right) \right\} dx\ dt$$

Using the same method, we find:

$$\begin{aligned}
&\psi'' - \frac{1}{c_p^2}\ddot{\psi} - \frac{1}{(2\lambda)^2}\psi = -\frac{T}{2}\frac{1}{(2\lambda)^2} \\
&U(x,t) = \frac{1}{2}\psi(x,t) \\
&V(x,t) = 0
\end{aligned}$$

In steady state, in the moving frame of reference, the intact zone upstream of the disturbance is the section $]L, +\infty]$. After performing the change of variable $x' = x - ct$, the stationary action is given by:

$$S(\psi) = b \cdot h \int_L^{+\infty} \left\{ \frac{1}{2}\rho \left(\frac{h}{2}(-c)\frac{\partial \psi}{\partial x'}\right)^2 + \frac{1}{2}\rho I \left(-c\frac{\partial \psi}{\partial x'}\right)^2 - \frac{1}{2}E\left(\frac{h}{2}\frac{\partial \psi}{\partial x'}\right)^2 - \frac{1}{2}\kappa G \psi^2 + \frac{\tau}{h}\left(\frac{h}{2}\psi\right) \right\} dx'$$

hence:

$$\begin{cases}
\left(1 - \frac{c^2}{c_p^2}\right)\psi^{(2)} - \frac{1}{(2\lambda)^2}\psi = -\frac{T}{4}\frac{1}{(2\lambda)^2} \\
U(x') = \frac{1}{2}\psi(x') \\
V(x') = 0
\end{cases}$$

*General solution of $\psi$ ($c < c_p$), non-divergent as $x$ goes to $\infty$:*

$$\psi(x') = K \exp\left(-\frac{x'}{2\lambda\sqrt{1-\frac{c^2}{c_p^2}}}\right) + \frac{T}{2}$$

By defining $\psi^+ = \psi(L)$,

$$\psi(x') = (\psi^+ - \tfrac{T}{2}) \exp\left(-\frac{(x'-L)}{2\lambda\sqrt{1-\frac{c^2}{c_p^2}}}\right) + \frac{T}{2}$$

## IV. Equations of motion of the collapsed section

The kinematic relationship now reads:

$$u\left(x, \frac{h}{2} + D_{WL}, t\right) = u(x,t) - \left(\frac{h}{2} + D_{WL}\right)\psi(x,t) = 0$$

and the other conditions are the same as for the intact section. In steady state, in the moving reference frame, the collapsed section is $]-\infty, 0[$. Over there, the stationary solution is deduced from the one on $]L, +\infty[$, by applying the non-divergence of $\psi$ at $-\infty$. We do not explicit it here, as it will not be used in this paper. Indeed, the knowledge of the solutions $(u, \psi)$ on $]-\infty, 0[$ and $]L, +\infty[$ is not necessary, as the central section (and its boundary conditions) prove to be sufficient to deduce the propagation speed of the perturbation.

## V. Assembling the solution on the collapsing central section in the Sub-Rayleigh speed range

In this section only, in order not to make the reading of the equations more arduous, we will refer to $\frac{d}{dx'}$ by using the notation $'$ previously attributed to $\frac{\partial}{\partial x}$. This is not be problematic, since we previously highlighted that $\frac{d}{dx'} = \frac{\partial}{\partial x}$ in steady state, allowing us to use the same notation for both derivatives.

### 1. Generic resolution of the assembly

As can be noticed above, *six* boundary conditions must be brought forth to account for the six unknowns on the central collapsing section (namely, the four integration constants $A, B, \mathcal{C}, D$, the speed $c$ and the touch-down distance $L$). To this end, in order to provide the most general results possible, we seek for a solution to the following set of generic boundary conditions:

$$V(x' = 0) = \frac{h_c}{h} \qquad V'(x' = L) = V'_L$$
$$V(x' = L) = 0 \qquad \psi'(x' = 0) = \psi'_0$$
$$V'(x' = 0) = 0 \qquad \psi'(x' = L) = \psi'_L$$

In the sub-Rayleigh regime, the set is therefore rewritten as:

$$\begin{cases} A_2 + D_2 = \frac{h_c}{h} & (22) \\ A_2\cos(\gamma) + B_2\sin(\gamma) + \frac{1}{2}\cdot\frac{c_s^2}{c^2}\cdot\frac{\Sigma_{\text{eff}}}{h^2}\cdot L^2 + C_2\cdot L + D_2 = 0 & (23) \\ \frac{\gamma}{L}\cdot B_2 + C_2 = 0 & (24) \\ \frac{\gamma}{L}\cdot(-A_2\sin(\gamma) + B_2\cos(\gamma)) + \frac{c_s^2}{c^2}\cdot\frac{\Sigma_{\text{eff}}}{h^2}\cdot L + C_2 = V'_L & (25) \\ \frac{c_s^2}{c^2}\cdot\frac{\Sigma_{\text{eff}}}{h} - h\frac{\gamma^2}{L^2}\cdot\left(1 - \frac{c^2}{c_s^2}\right)\cdot A_2 = \psi'_0 & (26) \\ \frac{c_s^2}{c^2}\cdot\frac{\Sigma_{\text{eff}}}{h} - h\frac{\gamma^2}{L^2}\cdot\left(1 - \frac{c^2}{c_s^2}\right)\cdot(A_2\cos(\gamma) + B_2\sin(\gamma)) = \psi'_L & (27) \\ L = \gamma\lambda\frac{c_s}{c}\cdot\Pi(c) & (28) \end{cases}$$

where we highlighted the boundary conditions. Note that we defined $\gamma = L/r(c)$ for simplicity, recalling that $r(c) = \lambda\frac{c_s}{c}\cdot\Pi(c)$, hence the additional equation (28). This system is solved by a series of substitutions and combinations:

(26) → $A_2 = \frac{\frac{c_s^2}{c^2}\frac{\Sigma_{eff}}{h^2} - \frac{\psi'_0}{h}}{\frac{\gamma^2}{L^2}\left(1-\frac{c^2}{c_s^2}\right)}$

(22) → $D_2 = \frac{h_c}{h} - A_2$

(24) → $C_2 L = -\gamma B_2$

(27)-(26) →
$$A_2\cos(\gamma) + B_2\sin(\gamma) = A + \frac{\psi'_0-\psi'_L}{h\frac{\gamma^2}{L^2}\left(1-\frac{c^2}{c_s^2}\right)}$$
$$\Leftrightarrow A_2(\cos(\gamma)-1) + B_2\sin(\gamma) = \frac{\psi'_0-\psi'_L}{h\frac{\gamma^2}{L^2}\left(1-\frac{c^2}{c_s^2}\right)}$$
$$\Leftrightarrow B_2 = \frac{1}{\sin(\gamma)}\left[\frac{\psi'_0-\psi'_L}{h\frac{\gamma^2}{L^2}\left(1-\frac{c^2}{c_s^2}\right)} + A_2(1-\cos(\gamma))\right] = \frac{1}{\sin(\gamma)}\frac{\psi'_0-\psi'_L}{h\frac{\gamma^2}{L^2}\left(1-\frac{c^2}{c_s^2}\right)} + \tan(\tfrac{\gamma}{2})\,A_2$$
$$\Leftrightarrow B_2 = \frac{1}{\sin(\gamma)}\frac{\psi'_0-\psi'_L}{h\frac{\gamma^2}{L^2}\left(1-\frac{c^2}{c_s^2}\right)} + \tan(\tfrac{\gamma}{2})\frac{\frac{c_s^2}{c^2}\frac{\Sigma_{eff}}{h^2} - \frac{\psi'_0}{h}}{\frac{\gamma^2}{L^2}\left(1-\frac{c^2}{c_s^2}\right)}$$

where we have used the well-known identity $\frac{1-\cos(\gamma)}{\sin(\gamma)} = \tan(\frac{\gamma}{2})$. Using all the previous results,

(23) →
$$A_2\cos(\gamma) + B_2\sin(\gamma) + \frac{c_s^2}{2c^2}\frac{\Sigma_{eff}}{h^2}L^2 + C_2 L + D_2 = 0$$
$$\Leftrightarrow A_2 + \frac{\psi'_0-\psi'_L}{h\frac{\gamma^2}{L^2}\left(1-\frac{c^2}{c_s^2}\right)} + \frac{c_s^2}{2c^2}\frac{\Sigma_{eff}}{h^2}L^2 - \gamma B_2 + \frac{h_c}{h} - A_2 = 0$$
$$\Leftrightarrow \frac{\psi'_0-\psi'_L}{h\frac{\gamma^2}{L^2}\left(1-\frac{c^2}{c_s^2}\right)} + L^2\frac{c_s^2}{2c^2}\frac{\Sigma_{eff}}{h^2} - \gamma\frac{1}{\sin(\gamma)}\frac{\psi'_0-\psi'_L}{h\frac{\gamma^2}{L^2}\left(1-\frac{c^2}{c_s^2}\right)} - \gamma\tan(\tfrac{\gamma}{2})\frac{\frac{c_s^2}{c^2}\frac{\Sigma_{eff}}{h^2} - \frac{\psi'_0}{h}}{\frac{\gamma^2}{L^2}\left(1-\frac{c^2}{c_s^2}\right)} + \frac{h_c}{h} = 0$$
$$\Leftrightarrow L^2\frac{c_s^2}{c^2}\frac{\Sigma_{eff}}{h^2}\left[\frac{1}{2}\left(1-\frac{c^2}{c_s^2}\right) + \frac{\psi'_0-\psi'_L}{\frac{c_s^2}{c^2}\frac{\Sigma_{eff}}{h}}\left[\frac{1}{\gamma^2} - \frac{1}{\gamma\sin(\gamma)}\right] - \frac{\tan(\frac{\gamma}{2})}{\gamma}\left(1 - \frac{\psi'_0}{\frac{c_s^2}{c^2}\frac{\Sigma_{eff}}{h}}\right)\right] = -\frac{h_c}{h}\left(1-\frac{c^2}{c_s^2}\right)$$

We now replace $L$ thanks to Equation (28) :

$$\frac{c_s^4}{c^4}\lambda^2\Pi^2(c)\gamma^2\frac{\Sigma_{eff}}{h^2}\left[\frac{1}{2}\left(1-\frac{c^2}{c_s^2}\right) + \frac{\psi'_0-\psi'_L}{\frac{c_s^2}{c^2}\frac{\Sigma_{eff}}{h}}\left[\frac{1}{\gamma^2} - \frac{1}{\gamma\sin(\gamma)}\right] - \frac{\tan(\frac{\gamma}{2})}{\gamma}\left(1 - \frac{\psi'_0}{\frac{c_s^2}{c^2}\frac{\Sigma_{eff}}{h}}\right)\right] = -\frac{h_c}{h}\left(1-\frac{c^2}{c_s^2}\right)$$
$$\frac{c_s^4}{c^4}\lambda^2\left(1-\frac{c^2}{c_p^2}\right)\frac{\Sigma_{eff}}{h^2}\left[\gamma\tan\left(\frac{\gamma}{2}\right)\left(1-\frac{\psi'_0}{\frac{c_s^2}{c^2}\frac{\Sigma_{eff}}{h}}\right) + \left[\frac{\gamma}{\sin(\gamma)} - 1\right]\frac{\psi'_0-\psi'_L}{\frac{c_s^2}{c^2}\frac{\Sigma_{eff}}{h}} - \frac{1}{2}\gamma^2\left(1-\frac{c^2}{c_s^2}\right)\right] = -\frac{h_c}{h} \tag{29}$$

Similarly, by reusing the previous equations (except Equation (23)), it can be shown:

(25)-(24) →
$$-\frac{\gamma}{L}\left[A_2\sin\gamma + B_2(1-\cos\gamma)\right] = V'_L - \frac{c_s^2}{c^2}\frac{\Sigma_{eff}}{h^2}L$$
$$\Leftrightarrow -\gamma\sin\gamma\left[A_2 + \frac{1}{\sin^2\gamma}\left[\frac{\psi'_0-\psi'_L}{h\frac{\gamma^2}{L^2}\left(1-\frac{c^2}{c_s^2}\right)}(1-\cos\gamma) + A(1-\cos\gamma)^2\right]\right] = V'_L L - \frac{c_s^2}{c^2}\frac{\Sigma_{eff}}{h^2}L^2$$
$$\Leftrightarrow -\gamma\underbrace{\sin\gamma\,(1+\tan^2\tfrac{\gamma}{2})}_{=2\tan(\frac{\gamma}{2})}A_2 = V'_L L - \frac{c_s^2}{c^2}\frac{\Sigma_{eff}}{h^2}L^2 + \gamma\tan\frac{\gamma}{2}\frac{\psi'_0-\psi'_L}{h\left(1-\frac{c^2}{c_s^2}\right)}\frac{L^2}{\gamma^2}$$
$$\Leftrightarrow -2\gamma\tan(\tfrac{\gamma}{2})\frac{\frac{c_s^2}{c^2}\frac{\Sigma_{eff}}{h^2} - \frac{\psi'_0}{h}}{\left(1-\frac{c^2}{c_s^2}\right)}\frac{L^2}{\gamma^2} = V'_L L - \frac{c_s^2}{c^2}\frac{\Sigma_{eff}}{h^2}L^2 + \gamma\tan\frac{\gamma}{2}\frac{\psi'_0-\psi'_L}{h\left(1-\frac{c^2}{c_s^2}\right)}\frac{L^2}{\gamma^2}$$
$$\Leftrightarrow -2\tan(\tfrac{\gamma}{2})\left[\frac{c_s^2}{c^2}\frac{\Sigma_{eff}}{h^2} - \frac{\psi'_0}{h}\right] = \frac{\gamma}{L}\left(1-\frac{c^2}{c_s^2}\right)V'_L \; - \frac{c_s^2}{c^2}\frac{\Sigma_{eff}}{h^2}\gamma\left(1-\frac{c^2}{c_s^2}\right) + \tan\frac{\gamma}{2}\left(\frac{\psi'_0-\psi'_L}{h}\right)$$
$$\Leftrightarrow \frac{c_s^2}{c^2}\frac{\Sigma_{eff}}{h^2}\left[\gamma\left(1-\frac{c^2}{c_s^2}\right) - 2\tan\frac{\gamma}{2}\right] = -\tan\frac{\gamma}{2}\;\left(\frac{\psi'_L+\psi'_0}{h}\right) + \left(1-\frac{c^2}{c_s^2}\right)\frac{\gamma}{L}V'_L$$

then by subtracting $L$ again through its expression (28):

$$\frac{c_s^2}{c^2}\frac{\Sigma_{eff}}{h^2}\left[\gamma\left(1-\frac{c^2}{c_s^2}\right)-2\tan\left(\frac{\gamma}{2}\right)\right]=-\tan\left(\frac{\gamma}{2}\right)\left(\frac{\psi_0'+\psi_L'}{h}\right)+\gamma\frac{\sqrt{1-\frac{c^2}{c_s^2}}}{\sqrt{1-\frac{c^2}{c_p^2}}}\frac{c}{c_s}\frac{1}{\lambda}V_L' \tag{30}$$

Equations (29) and (30), the *constitutive equation of the velocity* and *constitutive equation of* $\gamma$, respectively, consist of two coupled relations. Although it can't be achieved through analytical decoupling (given that it is very cumbersome with these generic boundary conditions), their numerical resolution allows to find both the speed $c$, and the touch-down distance $L$ through the knowledge of $\gamma$.

Several particular cases are studied in sub-sections (2) to (5).

## 2. Generalised "Heierli solution"

The quantity $\psi_n' \equiv \frac{c_s^2}{c^2}\frac{\Sigma_{eff}}{h}$ emerges as a natural value for the angular deformations $\psi_0'$ and $\psi_L'$. This is exploited by Heierli (2005) when writing, for a slab modelled as a plate on a flat ground, without fracture and compaction energies (eq. (6b) and (6c)):

$$v''(0)=0$$
$$v''(L)=\frac{g}{c^2}$$

In fact, in our notation system, these conditions are transcribed as follows:

$$V''(0)=0$$
$$v''(L)=\frac{\rho g h}{\rho h c^2}=\frac{\kappa G\Sigma_{\text{eff}}}{\rho h c^2}=\frac{c_s^2}{c^2}\frac{\Sigma_{eff}}{h}\Rightarrow V''(L)=\frac{c_s^2}{c^2}\frac{\Sigma_{eff}}{h^2}$$

and using the coupling equation (9), we find again:

$$\psi_L'=h\left(1-\frac{c^2}{c_s^2}\right)\frac{c_s^2}{c^2}\frac{\Sigma_{eff}}{h^2}+\frac{\Sigma_{eff}}{h}=\frac{c_s^2}{c^2}\frac{\Sigma_{eff}}{h}=\psi_n'$$
$$\psi_0'=\frac{\Sigma_{eff}}{h}$$

With this reasoning, we see that $\psi_n'$ is the angular deformation reflecting the free fall of the slab, which is natural in the absence of (i) energy contribution related to the fracturing and compaction of the weak layer and (ii) internal shear forces. Considering $\psi_L'=\frac{c_s^2}{c^2}\frac{\Sigma_{eff}}{h}$ , $\psi_0'=0$ and $V_L'=0$ thus restores a result near to the one that Heierli (2005) obtained for low speeds ($0<c\ll c_s$), but which remains valid for high speeds:

$$c=c_s\sqrt[4]{\frac{1}{2}\frac{E}{12(\kappa G)^2}\frac{h}{h_c}\left(\rho g h\cos(\theta)-w_{comp}^{vol}\right)\left(1-\frac{c^2}{c_p^2}\right)\left(1-\frac{c^2}{c_s^2}\right)} \tag{31}$$

$$L=\gamma\cdot\lambda\cdot\frac{c}{c_s}\sqrt{\left(1-\frac{c^2}{c_s^2}\right)\left(1-\frac{c^2}{c_p^2}\right)} \tag{32}$$

$$\frac{L}{r}=\gamma=\tan\left(\frac{\gamma}{2}\right)\approx 2.3311 \tag{33}$$

This set of solutions will be referred to as the "Generalised Heierli solution" in the rest of the paper, while (Eq. 7a) taken from Heierli (2005) will be called the "Heierli solution".

## 3. Considering fractures at the crack tip

On the other hand, if the weak layer is considered, the value of $\psi'_L$ ($\neq \psi'_n$ ) should be chosen by stress considerations at the edges, as shown below. To obtain the value of $V''$ at the edges, let us carry out the balance of the vertical forces at the end sections:

- At $x' = L$ :

$$\rho A \frac{\partial^2 V}{\partial t^2} = \underbrace{\rho A g}_{Gravity} + \underbrace{f_I^{lineic}}_{Fracturing} + \underbrace{f_{comp}}_{Compaction} + \underbrace{\frac{dQ}{dx}}_{Internal\ shear\ forces}$$

  where $f_{f,I}^{lin} = -w_{f,I}^{vol} \cdot D_{WL}$ is a force per unit length representing the resistance to fracturing, as it was shown that:

$$\mathcal{E}_{f,I} = w_{f,I}^{vol} \cdot b \cdot D_{WL} \cdot dx$$
$$\mathcal{E}_{f,I}^{lin} = w_{f,I}^{vol} \cdot b \cdot D_{WL}$$
$$f_{f,I}^{lin} = w_{f,I}^{vol} \cdot D_{WL} = w_{f,I}^{surf}$$

  Also, $\frac{dQ}{dx} = \kappa G A \left( h \frac{d^2V}{dx^2} - \frac{d\psi}{dx} \right)$. If the shear forces are neglected, and that no compaction is assumed at the edge,

$$\frac{\partial^2 V}{\partial t^2} \approx g - \frac{w_{f,I}^{surf}}{\rho A} \equiv g^*$$

  By recalling $\frac{d^2V}{dx'^2} = \frac{1}{c^2} \frac{\partial^2 V}{\partial t^2}$, the result reads $\frac{d^2V}{dx'^2} = \frac{g}{c^2} - \frac{w_{f,I}^{surf}}{\rho A c^2} = \frac{g^*}{c^2}$. Thus, we obtain a correction of the Heierli (2005) boundary condition, that now considers the amount of energy required to fracture the front of the collapse wave. The latter tends to slow down the free fall (everything happens as though an effective acceleration of gravity was virtually applied at the $x$' = $L$ border, lower than $g$).

- A similar result is obtained at $x' = 0$, except that the reaction of the substrate is now assumed to compensate all forces and cancel acceleration, so that:

$$\frac{d^2V}{dx'^2} = 0$$

Using Equation (9) and defining $\Sigma^* = \rho g^* h \cos(\theta) / \kappa G$, these conditions translate to:

$$\psi'(0) = \frac{\Sigma_{eff}}{h}$$
$$\psi'(L) = h \left(1 - \frac{c^2}{c_s^2}\right) \frac{\Sigma^*}{h^2} \frac{c_s^2}{c^2} + \frac{\Sigma_{eff}}{h} = \frac{\Sigma^*}{h} \frac{c_s^2}{c^2} + \frac{\Sigma_{eff} - \Sigma^*}{h}$$

Note that the existence of shear forces makes the previous reasoning invalid.

## 4. Solution for free edges

A system for which $\psi'_L = \psi'_0 = V'_L = 0$ (free edges at both ends with tangency at $x' = L$) does not provide propagative solutions in a permanent regime. Indeed, the constitutive relation of $\gamma$ becomes $\gamma\left(1 - \frac{c^2}{c_s^2}\right) = 2\tan(\frac{\gamma}{2})$, which then cancels the constitutive equation on the velocity, leading to $L = c = 0$. However, a solution with only one free edge (for instance, at $x' = 0$) would still be theoretically possible.

### 5. Free fall at both edges

Conversely, a constrained beam where gravity is imposed at both ends ($\psi'_L = \psi'_0 = \psi'_n$) with tangency at $x' = L$ will necessarily make transverse waves propagate at the shear S-waves speed ($c_s$), due to Equation (34). However, the boundary conditions themselves have to be reviewed, given that the condition $V'(0)$ cannot be satisfied anymore, as shown in Supplementary Material n°6.

## VI. Assembling the solution on the collapsing central section in the Supershear speed range

In this section, we show that a supershear purely flexural wave is impossible. As solving the problem with generic boundary conditions is tedious, we decide to prove it with the Heierli (2005) boundary conditions, which proved to be relatively relevant in the latter section. The set of boundary conditions then reads:

$$
\begin{cases}
A_1 + B_1 + D_1 = \dfrac{D_{WL}}{h} & (35)\\
A_1 \cdot \exp(\gamma) + B_1 \cdot \exp(-\gamma) + \dfrac{1}{2} \cdot \dfrac{c_s^2}{c^2} \cdot \dfrac{\Sigma_{\text{eff}}}{h^2} \cdot L^2 + C_1 \cdot L + D_1 = 0 & (36)\\
\dfrac{\gamma}{L} \cdot (A_1 - B_1) + C_1 = & (37)\\
\dfrac{\gamma}{L} \cdot (A_1 \cdot \exp(\gamma) - B_1 \cdot \exp(-\gamma)) + \dfrac{c_s^2}{c^2} \cdot \dfrac{\Sigma_{\text{eff}}}{h^2} \cdot L + C_1 = 0 & (38)\\
\dfrac{c_s^2}{c^2} \cdot \dfrac{\Sigma_{\text{eff}}}{h} + \gamma^2 \dfrac{h}{L^2} \cdot \left(1 - \dfrac{c^2}{c_s^2}\right) \cdot (A_1 + B_1) = 0 & (39)\\
\dfrac{c_s^2}{c^2} \cdot \dfrac{\Sigma_{\text{eff}}}{h} + \gamma^2 \dfrac{h}{L^2} \cdot \left(1 - \dfrac{c^2}{c_s^2}\right) \cdot (A_1 \cdot \exp(\gamma) + B_1 \cdot \exp(-\gamma)) = 0 & (40)\\
L \cdot c = \gamma \lambda c_s \cdot \Pi(c) & (41)
\end{cases}
$$

Contrarily to the *sub-Rayleigh* case, we do not need all those equations, as the incoherence comes from the constitutive equation of $\gamma$:

Equation (40) → $B_1 = -\frac{\exp(\gamma)}{\exp(-\gamma)} A_1 = -\exp(2\gamma)\, A_1$

Equations (40)-(39) → $A_1 \exp(\gamma) + B_1 \exp(-\gamma) = A_1 + B_1 + \frac{c_s^2}{c^2} \cdot \frac{\Sigma_{\text{eff}}}{h^2} \cdot \frac{L^2}{\gamma^2} \frac{1}{\left(1 - \frac{c^2}{c_s^2}\right)}$

$$\Leftrightarrow A_1(\exp(\gamma) - 1) + B_1(\exp(-\gamma) - 1) = \tfrac{c_s^2}{c^2} \cdot \tfrac{\Sigma_{\text{eff}}}{h^2} \cdot \tfrac{L^2}{\gamma^2} \tfrac{1}{\left(1 - \frac{c^2}{c_s^2}\right)}$$

$$\Leftrightarrow \tfrac{c_s^2}{c^2} \cdot \tfrac{\Sigma_{\text{eff}}}{h^2} \cdot \tfrac{L^2}{\gamma} = A_1 \gamma \left(1 - \tfrac{c^2}{c_s^2}\right)(\exp(\gamma) + 1)(\exp(\gamma) - 1)$$

We now combine the two previous results in Equations (37)-(38):

$$
\begin{aligned}
& A_1(1 - \exp(\gamma)) - B_1(1 - \exp(-\gamma)) = \tfrac{c_s^2}{c^2} \cdot \tfrac{\Sigma_{\text{eff}}}{h^2} \cdot \tfrac{L^2}{\gamma} \\
\Leftrightarrow & \left(1 - \exp(\gamma) + \exp(2\gamma)\left(1 - \exp(-\gamma)\right)\right) A_1 = \gamma A_1 \left(1 - \tfrac{c^2}{c_s^2}\right)(\exp(\gamma) + 1)(\exp(\gamma) - 1) \\
\Leftrightarrow & (\exp(\gamma) - 1)^2 = \gamma \left(1 - \tfrac{c^2}{c_s^2}\right)(\exp(\gamma) + 1)(\exp(\gamma) - 1) \\
\Leftrightarrow & (\exp(\gamma) - 1) = \gamma \left(1 - \tfrac{c^2}{c_s^2}\right)(\exp(\gamma) + 1) \\
\Leftrightarrow & \tfrac{\exp(\gamma) - 1}{\exp(\gamma) + 1} = \gamma \left(1 - \tfrac{c^2}{c_s^2}\right)
\end{aligned}
$$

We identifiy the trigonometrical formula $\frac{\exp(\gamma)-1}{\exp(\gamma)+1} = \tanh(\frac{\gamma}{2})$, from which stems:

$$\tanh\left(\frac{\gamma}{2}\right) = -\gamma \underbrace{\left(\frac{c^2}{c_s^2} - 1\right)}_{>0} \tag{42}$$

Given that $c > c_s$ in the *supershear* regime, Condition (42) is verified only for $\gamma = 0$, in which case the bending length and velocity are zero (non-propagative solution). In other words, this development points out that *the supershear regime in bending is unattainable.*

# Supplementary Material n°3 - Deriving the equations of motion in the supershear regime

This section develops a simple model for the supershear regime, based on Figure 12 of the paper.

The strategy adopted here is a generalisation of the one proposed by Trottet et al. (2022): propagation is dominated by longitudinal deformations $u$ after a brittle shear layer failure. In accordance with this, the transverse displacement $v$ is taken to be uniformly zero. Following the approximation of Trottet et al. (2022), the weak layer is thin and allows the stress within it to be considered uniform as a first approximation (internal shear neglected). The same approximation is made in the slab where the rotation angles $\psi$ are small: indeed, in the *sub-Rayleigh* propagation model, the weak layer was undeformable and imposed a total adhesion of the slab to the weak layer, resulting in zero displacement $u_x$ at their interface, and in large angular deformations in the slab. Here, however, the longitudinal displacement $u_x$ of the slab is allowed at the interface, either by coulombian sliding on the damaged weak layer or by elastic recovery of the displacement on the still intact weak layer. As a corollary, the angular displacement $\psi$ related to shear is neglected in the slab, which imposes uniformity of displacement in the slab: $u_x(x, y) \equiv u(x)$.

As a consequence of the above points, the weak layer constitutive law adopted in this section is illustrated the same as Trottet et al. (2022), namely:

$$\tau_{xy}(x) = \begin{cases} \tau_r = \tan(\phi) \cdot \rho g h \cos(\theta) & \forall x' \geq 0 \\ G_{WL} \dfrac{u(x)}{D_{WL}} = G_{WL} \gamma(x) & \forall x' < 0 \end{cases} \tag{43}$$

In this formula,

- $\phi$ is the friction angle linked to the coulombian friction coefficient by $\mu = \tan(\phi)$;
- $\tau_r$ denotes the constant residual friction of the weak layer once broken, which is a function of the collapse height (Trottet et al., 2022) assumed to be nil here;
- $G_{WL}$ refers to the shear modulus of the weak layer;
- Finally, as there is no bending anymore (thus, no bending length), the only point of interest is the place where breakage takes place, which we choose to define at $x' = 0$ such that $u(0) = u_p = \frac{\tau_p^2}{G_{WL}}$ (Figure 1).

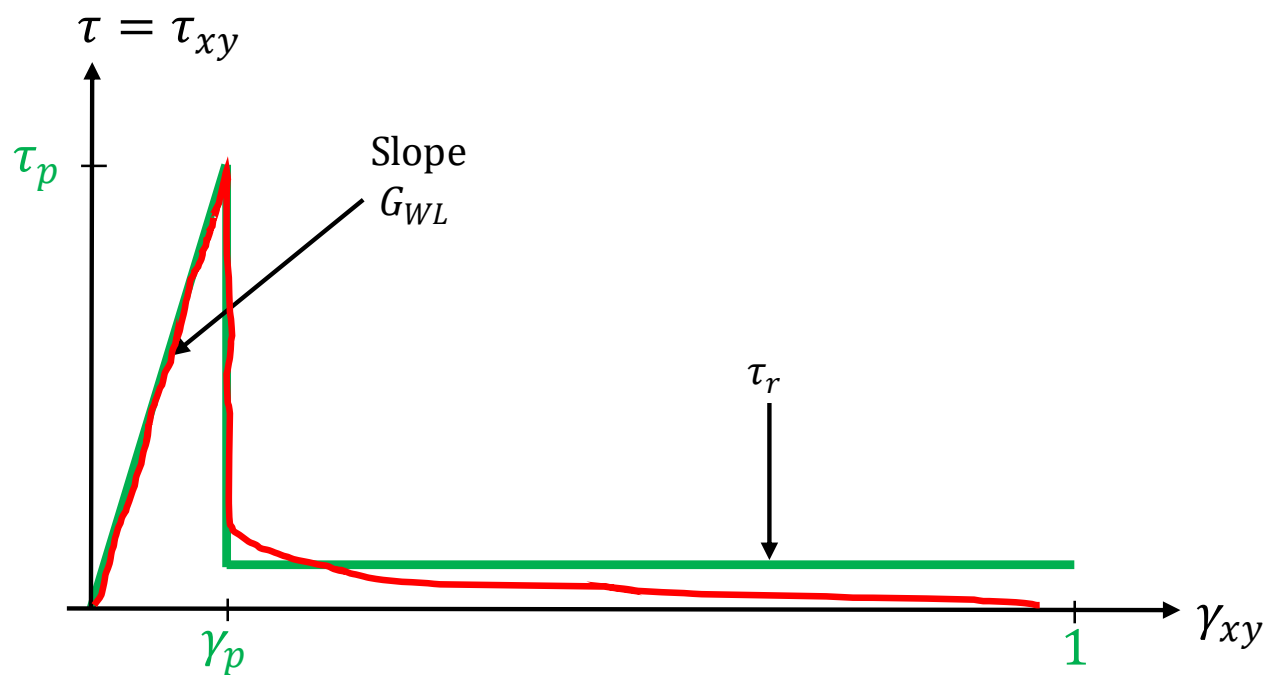

Figure 1 – Diagram of the behaviour of the weak layer in shear. (*Red*) Realistic constitutive law. (*Green*) Simplified brittle model.

On the $]-\infty, 0[$ section, the restoring force in the weak layer translates into a potential energy in the slab overlying the still-intact portion of the weak layer:

$$\delta\mathbb{E} = bD_{WL}dx \cdot \int_{WL} \tau d\gamma = bD_{WL}dx \cdot \int_{WL} G_{WL}\gamma d\gamma = -\frac{1}{2}G_{WL}\gamma^2 \cdot D_{WL} \cdot bdx = -\frac{1}{2}G_{WL}\frac{u^2}{D_{WL}h} \cdot Adx \tag{44}$$

## 1. Solving the problem on the intact section $]-\infty, \mathbf{0}[$

The action functional of the slab on the intact weak layer section reads

$$S(u,t) = A \cdot \int_0^t \left\{ \int_{x_1}^{x_2} \left\{ \frac{1}{2}\rho\dot{u}^2 - \frac{1}{2}Eu'^2 + \frac{\tau_g}{h}u - \frac{1}{2}\frac{G_{WL}}{D_{WL}h}u^2 \right\} dx - \frac{1}{A}\mathcal{N}_{x_1,x_2}u \right\} dt$$

where $\mathcal{N}_{x_1,x_2}$ are the external normal forces applied at both ends.

The motion for the slab is given by:

$$Eu'' - \rho\ddot{u} - \frac{G_{WL}}{D_{WL}^2}u = -\frac{\tau_g}{h}$$

In steady state,

$$\left(1 - \frac{c^2}{c_p^2}\right)u^{(2)} - \frac{G_{WL}}{ED_{WL}h}u = -\frac{\tau_g}{Eh} \tag{45}$$

Since $c = c_p$ is not acceptable on the other section (see Equation (48)), Equation (45) becomes:

$$u^{(2)} - \frac{1}{\lambda^2}u = -\frac{\tau_g}{Eh\left(1 - \frac{c^2}{c_p^2}\right)}$$

with $\lambda(c)$ is a characteristic stress relaxation distance:

$$\lambda(c) = \sqrt{\frac{EhD_{WL}}{G_{WL}}\left(1 - \frac{c^2}{c_p^2}\right)} = \lambda_{stat} \cdot \sqrt{1 - \frac{c^2}{c_p^2}} \tag{46}$$

The solution satisfying the kinematic and stress boundary conditions ($u(0) = u_p$ and $u'(-\infty) \to 0$) reads:

$$u(x' < 0) = \left(u_p - \frac{\tau_g D_{WL}}{G_{WL}}\right)e^{\frac{x'}{\lambda(c)}} + \frac{\tau_g D_{WL}}{G_{WL}}$$

$$u^{(1)}(x' < 0) = \frac{u_p - \frac{\tau_g}{G_{WL}}D_{WL}}{\lambda(c)}e^{\frac{x'}{\lambda(c)}}$$

Thus, in particular,

$$u^{(2)}(x' < 0) = \frac{u_p - \frac{\tau_g}{G_{WL}}D_{WL}}{\lambda_{stat}^2\left(1 - \frac{c^2}{c_p^2}\right)}e^{\frac{x'}{\lambda(c)}} \tag{47}$$

## 2. Solving the problem on the sliding section $[\mathbf{0}, +\infty[$

The action functional of the slab on the damaged weak layer section is

$$S(u,t) = A \cdot \int_0^t \left\{ \int_{x_1}^{x_2} \left\{ \frac{1}{2}\rho\dot{u}^2 - \frac{1}{2}Eu'^2 + \frac{\tau_g - \tau_r}{h}u \right\} dx - \frac{1}{A}Q_{x_1,x_2}u \right\} dt$$

where $\tau_g \cdot h = \rho Ag\sin(\theta)$ and $\tau_r \cdot h = \rho Ag\tan(\phi)\cos(\theta)$, hence the equations of unsteady motion

$$Eu'' - \rho\ddot{u} = -\frac{\tau_g - \tau_r}{h}$$

and of steady motion

$$\left(1-\frac{c^2}{c_p^2}\right)u^{(2)}=-\frac{\tau_g-\tau_r}{Eh} \tag{48}$$

This equation underlines that propagating at $c=c_p$ simultaneously with a finite second derivative $U^{(2)}<\infty$ is not acceptable for this problem. One could then write:

$$u^{(2)}=-\frac{\tau_g-\tau_r}{Eh\left(1-\frac{c^2}{c_p^2}\right)} \tag{49}$$

whose general solution is

$$u(x')=-\frac{\tau_g-\tau_r}{2Eh\left(1-\frac{c^2}{c_p^2}\right)}x'^2+A\cdot x'+B$$

Here, $A$ et $B$ are integration constants determined by the conditions for connection at $x'=0$:

$$u(0^+)=u(0^-)=u_p \Leftrightarrow B=u_p$$
$$u'(0^+)=u'(0^-) \Leftrightarrow A=\frac{u_p-\frac{\tau_g}{G_{WL}}D_{WL}}{\lambda(c)}$$

## 3. Assembling the solution at the fracture point $x'=0$

Attempting to equalise the acceleration to the left and right of the crack, namely Equations (47) and (49), leads to an obvious contradiction, since the speed vanishes from both terms and results in an absurd equality between the remaining constants. Therefore, we expect a discontinuity of the acceleration at the crack tip. Once again, we seem to be bound to postulate an arbitrarily chosen acceleration at the crack tip, which should be selected in consideration of the available experimental data. For this purpose, we generalize the approach applied by Heierli (2005) to the bending case: similarly to its "free fall" in the *sub-Rayleigh* regime, the slab can be considered to "slide freely downwards" on the inclined damaged weak layer with a residual friction $\tau_r$, as soon as the fracture has released the slab from the restoring force of the weak layer. As a result, an acceleration proportional to gravity - residual friction subtracted - can be imposed at the tip. Let $\alpha$ denote the proportionality factor:

$$\frac{\partial^2 u}{\partial t^2}=\alpha\frac{\tau_g-\tau_r}{\rho}$$

hence $\frac{d^2u}{dx'^2}(0)=\frac{1}{c^2}\frac{\partial^2 u}{\partial t^2}(0)=\alpha\frac{\tau_g-\tau_r}{\rho h c^2}$. Thus, by equalising $u^{(2)}(0^+)=u^{(2)}(0)$, one gets

$$\frac{\tau_g-\tau_r}{Eh\left(1-\frac{c^2}{c_p^2}\right)}=\alpha\frac{\tau_g-\tau_r}{\rho h c^2}$$

and acknowledging $\frac{E}{\rho}=c_p$,

$$\left(1-\frac{c^2}{c_p^2}\right)=\frac{1}{\alpha}\frac{c^2}{c_p^2}$$
$$\Leftrightarrow \qquad c=c_p\sqrt{\frac{1}{1+\frac{1}{\alpha}}}$$

When $\alpha=1$, in particular, $c=\frac{c_p}{\sqrt{2}}$.

Several comments can be made: (i) firstly, in this model, the speed is not affected by any property either of the weak layer or of the slab, which could seem awkward at first glance. Nevertheless, it remains a good approximation regarding the available data, in which the physical parameters seem to affect the speed of the convergence towards the permanent regime, but not the permanent regime itself (Trottet et al., 2022); (ii) besides, in this approach, the form of the acceleration at the crack tip determines the resulting speed, so that once again, the physics of the problem is wholly contained in the acceleration at the *crack tip*; (iii) lastly, in this model, $c_p$ is the upper bound of the speeds, recovered only if the acceleration (through the control parameter $\alpha$) goes to infinity at the crack tip.

Bringing meaning to the factor $\alpha$ can be achieved by generalising the previous results to account for the release process at $x' = 0$. Indeed, when the fracture in the weak layer occurs, the slab can be expected to instantly recover the energy previously stored in the weak layer as a restoring force. This translates to an excess acceleration through a "boost" force $\overrightarrow{f_b} = f_b^{lin} \cdot dx\ \overrightarrow{e_x}$ where the linear force $f_b^{lin}$ is associated with a stress $\tau_b$ such that $f_b^{lin} = \tau_b \cdot b$. By balancing forces at $x' = 0$, *as soon as the internal elastic restoring force at $x' = 0$ is neglected in the slab*,

$$\rho b h \frac{\partial^2 u}{\partial t^2}(0) = b(\tau_g - \tau_r) + f_b^{lin} + \underbrace{\frac{d\mathcal{N}}{dx'}(0)}_{neglected} = b(\tau_g - \tau_r + \tau_b)$$

$$\frac{d^2 u}{dx'^2}(0) = \frac{1}{c^2}\frac{\partial^2 u}{\partial t^2}(0) = \frac{\tau_g - \tau_r + \tau_b}{\rho h c^2} = \alpha \frac{\tau_g - \tau_r}{\rho h c^2}$$

with $\alpha = 1 + \frac{\tau_b}{\tau_g - \tau_r}$. The continuity of the second derivative at $x' = 0$ becomes:

$$\frac{\tau_g - \tau_r}{Eh\left(1 - \frac{c^2}{c_p^2}\right)} = \frac{\tau_g - \tau_r}{\rho h c^2} + \frac{\tau_b}{\rho h c^2}$$

$$(\tau_g - \tau_r)\left(\frac{c^2}{c_p^2 - c^2} - 1\right) = \tau_b$$

$$c^2 = \left(1 + \frac{\tau_b}{\tau_g - \tau_r}\right)(c_p^2 - c^2)$$

$$c^2\left(2 + \frac{\tau_b}{\tau_g - \tau_r}\right) = \left(1 + \frac{\tau_b}{\tau_g - \tau_r}\right)c_p^2$$

$$c = \frac{c_p}{\sqrt{2}}\sqrt{\frac{\left(1 + \frac{\tau_b}{\tau_g - \tau_r}\right)}{\left(1 + \frac{1}{2}\frac{\tau_b}{\tau_g - \tau_r}\right)}}$$

Considering finally, with Eq.(43), that the "boost stress" is equal to the weak layer restoring stress at $x' = 0$, one can get an expression for $\tau_b$:

$$\tau_b \approx \tau_p = \frac{1}{bdx}\left(\frac{\delta\mathbb{E}}{du}\right)(u = u_p) = G_{WL}\frac{u_p}{D_{WL}}$$

In this case,

$$c = \frac{c_p}{\sqrt{2}}\sqrt{\frac{\left(1 + G_{WL}\frac{u_p}{D_{WL}}\frac{1}{\tau_g - \tau_r}\right)}{\left(1 + \frac{1}{2}G_{WL}\frac{u_p}{D_{WL}}\frac{1}{\tau_g - \tau_r}\right)}} \tag{50}$$

## Supplementary Material n°4 - Alternative modelling of the weak layer with a brittle failure in the Sub-Rayleigh regime

This section takes a 'strength-of-materials' perspective in the sub-Rayleigh regime, where transverse strains $v$ surpass longitudinal strains $u$. Since the compression mode (-I) seems predominant, we adopt the simplified model of a brittle weak layer behaviour in compression, established in

Figure 2. Since the compression mode (-I) is predominant, we adopt a simplified model of a brittle behaviour for the weak layer in compression, as can be seen in

Figure 2. Let $\sigma$ be the reaction-stress of the weak layer on the slab, and $\epsilon = v(x,t)/D_{WL}$ the compressive strain of the weak layer. The failure is brittle: the yield strength is equal to the ultimate strength of the material, and is denoted $\sigma_p = \sigma_{peak}$. The corresponding elongation at break is $\epsilon_{peak} = \epsilon_p = v_p/D_{WL}$. The model is then written as:

$$\sigma = \begin{cases} E_{WL} \cdot \epsilon & \textit{while } \epsilon < \epsilon_p \\ 0 & \textit{otherwise} \end{cases}$$

On a plane cross-section of the beam of length $dx$ and thickness $b$, the corresponding energy is:

$$\mathbb{E}_{compression} = \begin{cases} \frac{1}{2}\frac{E_{WL}}{D_{WL}} \cdot v^2(x,t) \cdot b \cdot dx & \textit{while } v < v_p \\ 0 & \textit{otherwise} \end{cases}$$

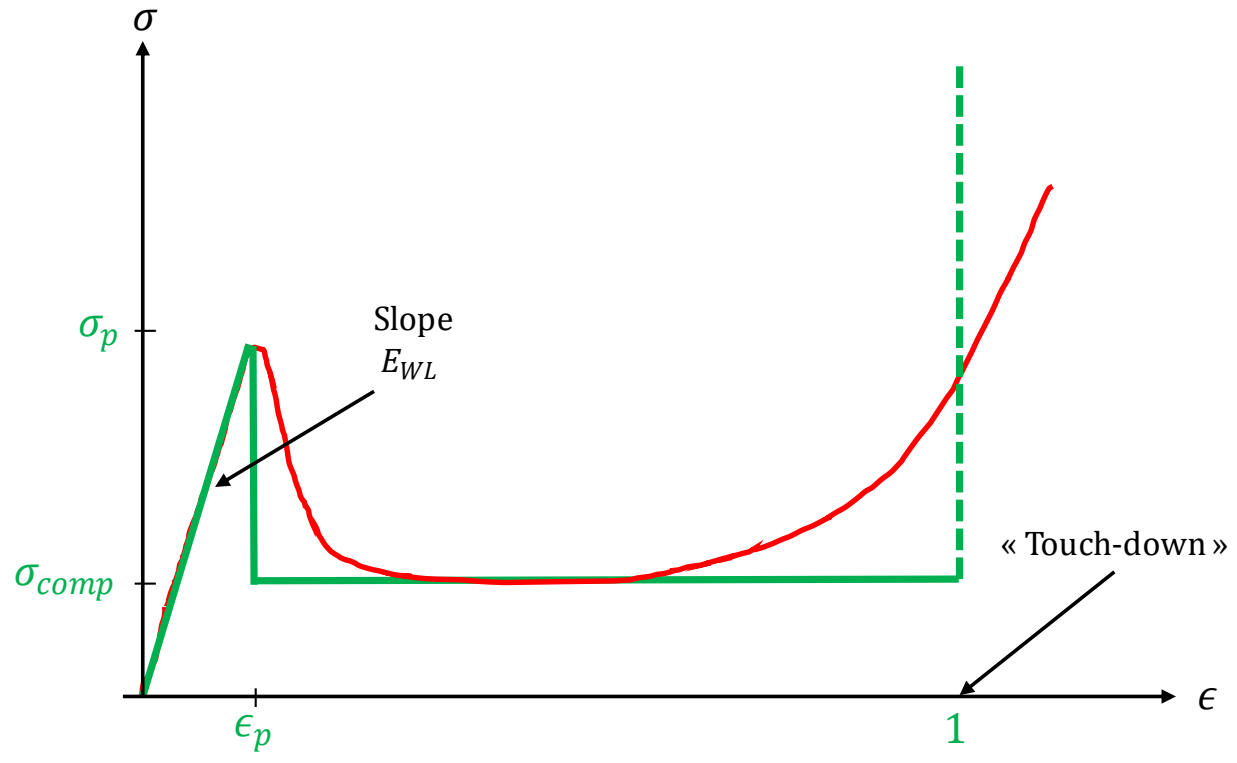

Figure 2 – Schematic of the weak layer compressive constitutive law. (*Red*) Realistic constitutive law. (*Green*) Simplified brittle model. *Reading*: the "touch-down" refers to the point where the weak layer reaches its maximum compression and cannot be compacted anymore; thus, the slab rests on an undeformable substrate, without any further deformation possible. Once brittle failure is reached, a constant stress $\sigma_{comp}$ is applied to the slab by the weak layer to resist to compaction.

In the moving reference frame linked to the disturbance in uniform translation at speed $c$, the slab is immobile. The position $x'_p = L$ of the brittle failure in this frame of reference is therefore also stationary, although unknown. The slab can then be divided into three sections with distinct stresses, as shown in Figure 3.

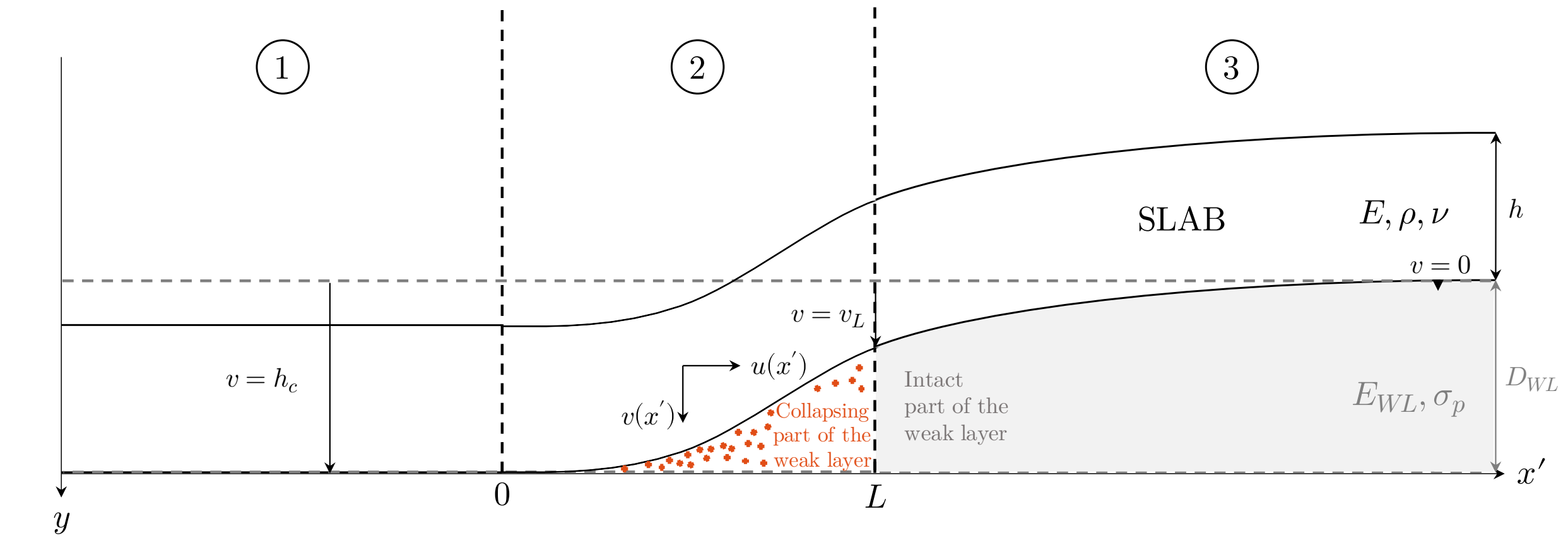

FIGURE 3 – Division of the slab into three sections in the moving reference frame in the *sub-Rayleigh* regime. (*Section 1*) From $-\infty$ to $x' = 0$, the beam rests uniformly on the substrate. (*Section 2*) From $x' = 0$ to $x' = L$, the weak layer collapses progressively according to the adopted constitutive law, under the effect of the falling slab which undergoes a compaction reaction of the weak layer during its bending (resistance energy $w^{vol}_{comp}$). (*Section 3*) From $x' = L$ to $+\infty$, the beam is supported by the weak layer elastically compacted by the weight of the slab. Brittle failure occurs at $L$. Note: the diagram is not to scale.

Following the same strategy as for the fracture mechanics model, the dimensionless differential equations of motion on the three sections are obtained below. For convenience, we define $\alpha = \frac{E_{WL}}{\kappa G D_{WL}}$ and $\beta = \frac{1}{\lambda^2} = \frac{\kappa G}{EI}$, so that $[\alpha] = [\beta] = L^{-2}$.

| Section 1 | Section 2 | Section 3 (we highlight the additional elastic term) |
|---|---|---|
| $U(x,t) = \frac{1}{2}\psi(x,t)$ | $U'' - \frac{1}{c_p^2}\ddot{U} = -\frac{\kappa G}{Eh^2}T$ | $U'' - \frac{1}{c_p^2}\ddot{U} = -\frac{\kappa G}{Eh^2}T$ |
| $V(x,t) = 0$ | $V'' - \frac{1}{c_s^2}\ddot{V} - \frac{1}{h}\psi = -\frac{\Sigma}{h^2}$ | $V'' - \frac{1}{c_s^2}\ddot{V} - \overbrace{\frac{E_{WL}}{\kappa G D_{WL}}}^{\alpha} V - \frac{1}{h}\psi = -\frac{\Sigma}{h^2}$ |
| $\psi'' - \frac{1}{c_p^2}\ddot{\psi} - \frac{1}{(2\lambda)^2}\psi = -\frac{T}{2}\frac{1}{12}\frac{1}{(2\lambda)^2}$ | $\psi'' - \frac{1}{c_p^2}\ddot{\psi} - \frac{\kappa G}{EI}\psi + \frac{\kappa G}{EI}hV' = 0$ | $\psi'' - \frac{1}{c_p^2}\ddot{\psi} - \underbrace{\frac{\kappa G}{EI}}_{\beta}\psi + \frac{\kappa G}{EI}hV' = 0$ |

In steady state, defining $x' = x - ct$, one obtains:

| Section 1 | Section 2 | Section 3 |
|---|---|---|
| $U(x,t) = \frac{1}{2}\psi(x,t)$ | $\left(1 - \frac{c^2}{c_p^2}\right) U^{(2)} = -\frac{\beta}{12}T$ | $\left(1 - \frac{c^2}{c_p^2}\right) U^{(2)} = -\frac{\beta}{12}T$ |
| $V(x,t) = 0$ | $\left(1 - \frac{c^2}{c_s^2}\right) V^{(2)} - \frac{1}{h}\psi = -\frac{\Sigma}{h^2}$ | $\left(1 - \frac{c^2}{c_s^2}\right) V^{(2)} - \alpha V - \frac{1}{h}\psi = -\frac{\Sigma}{h^2}$ |
| $\left(1 - \frac{c^2}{c_p^2}\right) \psi^{(2)} - \frac{\beta}{4}\psi = -\frac{1}{24}\frac{\beta}{4}T$ | $\left(1 - \frac{c^2}{c_p^2}\right) \psi^{(2)} - \beta\psi + \beta h V^{(1)} = 0$ | $\left(1 - \frac{c^2}{c_p^2}\right) \psi^{(2)} - \beta\psi + \beta h V^{(1)} = 0$ |

hence the uncoupled equations on $V$ alone:

– Section n°2:

$$\left(1 - \frac{c^2}{c_s^2}\right)\left(1 - \frac{c^2}{c_p^2}\right) V^{(4)} + \beta \frac{c^2}{c_s^2} V^{(2)} = \frac{\beta}{h^2}\Sigma \tag{51}$$

– Section n°3:

$$\left(1 - \frac{c^2}{c_s^2}\right)\left(1 - \frac{c^2}{c_p^2}\right) V^{(4)} + \left(\frac{c^2}{c_s^2}\beta - \alpha\left(1 - \frac{c^2}{c_p^2}\right)\right) V^{(2)} + \alpha\beta V = \frac{\beta}{h^2}\Sigma \tag{52}$$

In equation (52), the term $\alpha$ is related to the elastic contribution of the weak layer: considering $\alpha = 0$ allows to recover the case of Section n°2. The energy to be released to compress the weak layer (*via* $\alpha$) comes from the deformation energy of the beam (*via* $\beta$). The weak layer is therefore no longer in direct competition with gravity, as was the case in the fault mechanics model (it remains so indirectly, since the potential energy of gravity generates the deformation energy of the slab which will eventually be transferred into dissipated energy of deformation of the weak layer).

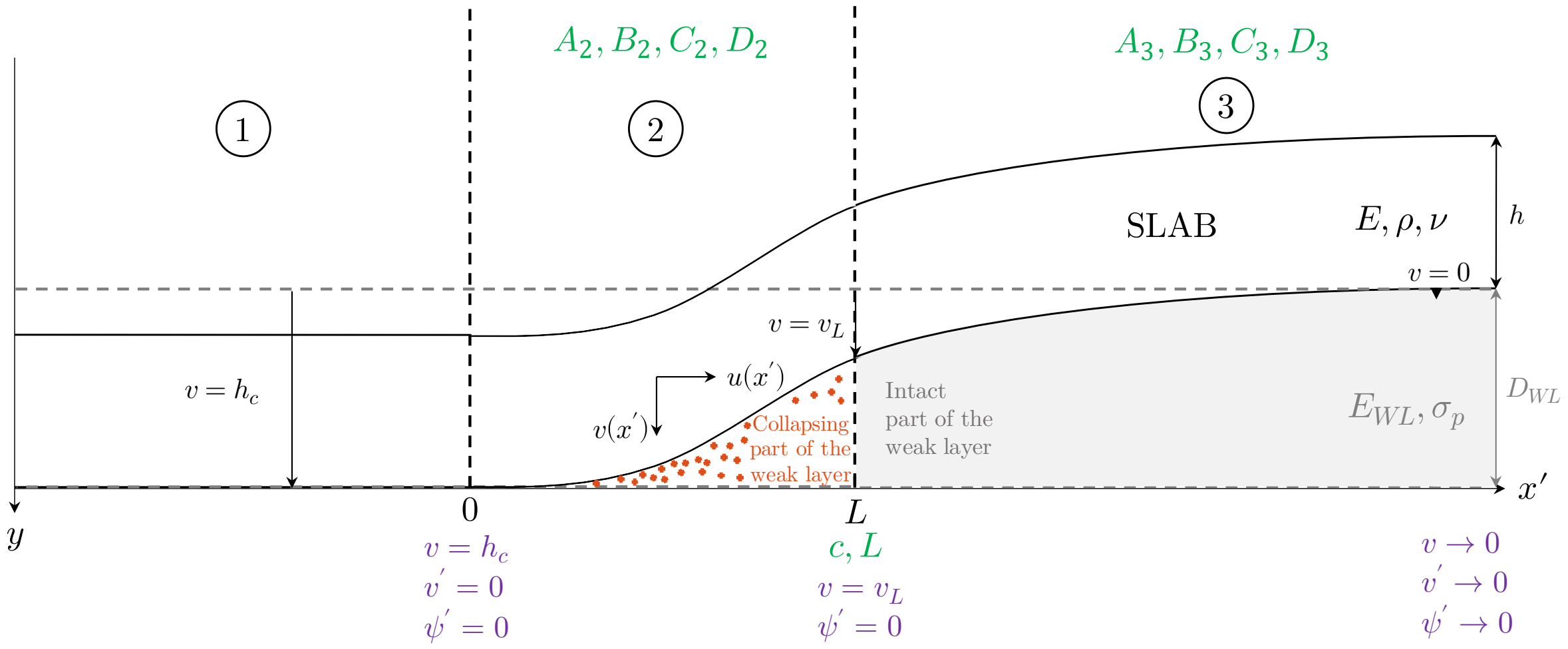

FIGURE 4 – Boundary conditions applied to the sections (*in purple*) and unknowns of the problem (*in green*). *Note*: the diagram is not to scale.

The selected boundary conditions are described below and summarised in Figure 4:

– Kinematic conditions:

$$\left.\begin{array}{r} v(x'=0) = 0 \\ v'(x'=0) = 0 \end{array}\right\} \quad \text{CONTINUITY of } v \text{ and } v' \text{ at } x'=0$$

$$\left.\begin{array}{r} v(x'=L) = v_L \\ v'(L^-) = v'(L^+) \end{array}\right\} \quad \text{CONTINUITY of } v \text{ and } v' \text{ at } x'=L$$

$$\left. v'(x') \xrightarrow[x'\to\infty]{} 0 \right\} \quad \text{LOCALITY of the action of the disturbance}$$

– Stress conditions:

$$\left. \psi'(x') \xrightarrow[x'\to\infty]{} 0 \right\} \quad \text{Hamilton's principle: } \psi' = 0 \text{ at all free edges}$$

Looking for the four independent solutions in the form $V_i(x') = e^{kx}$, the characteristic polynomial of equation (52) reads:

$$\left(1 - \frac{c^2}{c_s^2}\right)\left(1 - \frac{c^2}{c_p^2}\right) k^4 + \left(\frac{c^2}{c_s^2}\beta - \alpha\left(1 - \frac{c^2}{c_p^2}\right)\right) k^2 + \alpha\beta = 0$$

This is a bi-squared equation, which can be reduced to an equation of degree 2 by defining $K = k^2$, and whose determinant is:

$$\Delta = \left(\frac{c^2}{c_s^2}\beta - \alpha\left(1 - \frac{c^2}{c_p^2}\right)\right)^2 - 4 \cdot \alpha\beta \cdot \left(1 - \frac{c^2}{c_s^2}\right)\left(1 - \frac{c^2}{c_p^2}\right)$$

$$\Delta = \alpha^2\left(1 - c^2\left(\frac{\beta/\alpha}{c_s^2} + \frac{1}{c_p^2}\right)\right)^2 - 4 \cdot \alpha\beta \cdot \left(1 - \frac{c^2}{c_s^2}\right)\left(1 - \frac{c^2}{c_p^2}\right)$$

The determinant is a fourth-order polynomial in the speed $c$. In the absence of information on the orders of magnitude of $\alpha, \beta, c$, the generic roots cannot be obtained simply: approximations turn out to be necessary.

## 1. Low-speed limit

For instance, let us consider the low-velocity limit first ($c \ll c_s < c_p$):

$$\Delta \approx \alpha^2 \left(1 - 2\left(\frac{\frac{\beta}{\alpha}}{c_s^2} + \frac{1}{c_p^2}\right) \cdot c^2\right) - 4 \cdot \alpha\beta \cdot \left(1 - \frac{c^2}{c_s^2}\right)$$
$$\Delta \approx \alpha^2 - 4 \cdot \alpha\beta - \left(2\left(\frac{\alpha\beta}{c_s^2} + \frac{\alpha^2}{c_p^2}\right) + \frac{4\alpha\beta}{c_s^2}\right) c^2$$
$$\Delta \approx \alpha(\alpha - 4\beta) - 2\alpha\left(\frac{3\beta}{c_s^2} + \frac{\alpha}{c_p^2}\right) c^2$$

Please notice that $\Delta \approx \alpha^2 - 4 \cdot \alpha\beta$ is the determinant found by (Rosendahl & Weißgraeber, 2020) when considering an equivalent static system; in low speed dynamics, a correction term is therefore added.

$$\Delta = 0 \Leftrightarrow \frac{\alpha - 4\beta}{2\left(\frac{3\beta}{c_s^2} + \frac{\alpha}{c_p^2}\right)} = c_c^2 \Leftrightarrow c_c = c_s \cdot \sqrt{\frac{\alpha - 4\beta}{2\left(3\beta + \alpha\frac{c_s^2}{c_p^2}\right)}} \underset{c_s \approx \frac{c_p}{2}}{\approx} c_s \cdot \sqrt{\frac{\alpha - 4\beta}{2\left(3\beta + \frac{\alpha}{4}\right)}}$$

Two regimes can be distinguished with respect to the values of ($\alpha$,$\beta$):

- <u>$\alpha > 4\beta$ *and* $c < c_c$</u>: $\Delta > 0 \Rightarrow K = \frac{\alpha}{2\left(1-\frac{c^2}{c_s^2}\right)}\left(1 - c^2\left(\frac{\frac{\beta}{\alpha}}{c_s^2} + \frac{1}{c_p^2}\right) \pm \sqrt{\left(1 - \frac{4\beta}{\alpha}\right) - 2\left(\frac{\frac{3\beta}{\alpha}}{c_s^2} + \frac{1}{c_p^2}\right)c^2}\right) > 0$

  $$\Rightarrow k_{1,2} = \sqrt{\frac{\alpha}{2\left(1-\frac{c^2}{c_s^2}\right)}\left(1 - c^2\left(\frac{\frac{\beta}{\alpha}}{c_s^2} + \frac{1}{c_p^2}\right) \pm \sqrt{\left(1 - \frac{4\beta}{\alpha}\right) - 2\left(\frac{\frac{3\beta}{\alpha}}{c_s^2} + \frac{1}{c_p^2}\right)c^2}\right)}$$

  hence $V(x') = Ae^{k_1x'} + Be^{-k_1x'} + Ce^{k_2x'} + De^{-k_2x'} + \frac{1}{\alpha h^2}\Sigma$.
  Since $V(x')$ remains bounded around $+\infty$, $V(x') = Be^{-k_1x'} + De^{-k_2x'} + \frac{1}{\alpha h^2}\Sigma$.

- <u>$\alpha < 4\beta$</u> *or* <u>$\alpha > 4\beta$ *and* $c > c_c$</u>: $\Delta < 0 \quad \Rightarrow K = \frac{\alpha}{2\left(1-\frac{c^2}{c_s^2}\right)}\left(1 - c^2\left(\frac{\frac{\beta}{\alpha}}{c_s^2} + \frac{1}{c_p^2}\right) \pm i\sqrt{|\Delta|}\right)$

  $$\Rightarrow k_{1,2} = \left(\frac{\alpha}{2\left(1-\frac{c^2}{c_s^2}\right)}\left(1 - c^2\left(\frac{\frac{\beta}{\alpha}}{c_s^2} + \frac{1}{c_p^2}\right) \pm i\sqrt{|\Delta|}\right)\right)^{1/2} \in \mathbb{C}$$

  hence $V^*(x') = A^*e^{k_1x'} + B^*e^{-k_1x'} + C^*e^{k_2x'} + D^*e^{-k_2x'} + \frac{1}{\alpha h^2}\Sigma$.
  Given that $V^*$ is complex here, the only relevant profile $V$ is its real part: $V(x) = Re(V^*)$. Thus, by writing $k_1 = (a + ib)^{1/2} = r^{1/2}e^{i\theta/2}$ and $k_2 = (a - ib)^{1/2} = r^{1/2}e^{-i\theta/2}$, it is clear that:

  $$k_1 = k_3 + ik_4$$
  $$k_2 = k_3 - ik_4$$

  with $k_3 = r^{1/2}\cos(\frac{\theta}{2})$ and $k_4 = r^{1/2}\sin(\frac{\theta}{2})$. One then shows that:

  $$V(x') = e^{-k_3x'}(A\cos(k_4x') + B\sin(k_4x')) + e^{k_3x'}(C\cos(k_4x') + D\sin(k_4x'))$$

  Since $V(x')$ remains bounded around $+\infty$, $V(x') = e^{-k_3x'}(A\cos(k_4x') + B\sin(k_4x')) + \frac{1}{\alpha h^2}\Sigma$.

We therefore report the existence of a critical speed $c_c$ which distinguishes two propagation regimes: a decreasing exponential one, on the one hand, and an oscillating damped one, on the other. In the general case where $c \in ]0, c_s[$, the solutions can be of three types: decreasing exponential, damped oscillator or harmonic oscillator. As the last regime is not physically admissible, only the first two should be obtained with appropriate slab and weak layer parameters.

In any case, this low-speed approximation shows that a damped oscillating solution is admissible: this reasoning could therefore pave the road towards an explanation of the oscillations observed in the numerical and experimental acceleration curves (see an example in Figures B5 and 3.7 in Bobillier (2022)).

## 2. Rescaling of the equations of motion

Let $X$' denote $\frac{x'}{L}$. By defining $\frac{d^k}{(dX')^k} = \cdot^{[k]}$, the complete rescaling of the equations gives:

$$\left(1 - \frac{c^2}{c_s^2}\right)\left(1 - \frac{c^2}{c_p^2}\right) \cdot V^{[4]} + \left(\frac{c^2}{c_s^2}\beta - \alpha\left(1 - \frac{c^2}{c_p^2}\right)\right) \cdot L^2 \cdot V^{[2]} + \alpha\beta \cdot L^4 \cdot V = \frac{\beta}{h^2}\mathrm{L}^4 \cdot \Sigma$$

Typical numerical values are considered for the properties of the layers:

$$\alpha = \frac{E_{WL}}{hD_{WL}\kappa G} \sim \frac{2 \cdot E_{WL}}{hD_{WL}\kappa E} \sim \frac{2 \cdot 0.15}{1 \cdot 10.10^{-2} \cdot 1 \cdot 4} \sim 10^1$$

$$\beta = \frac{\kappa G}{EI} \sim \frac{\kappa E \cdot 12}{2E \cdot h^2} \sim \frac{6}{h^2} \sim \frac{6}{1^2} \sim 10^1$$

Unfortunately, these two terms seem to be of the same order of magnitude, which makes it irrelevant *a priori* to neglect one of them with respect to the other. Finally, Bergfeld et al. (2022) report typical experimental flexural lengths $L$ between 1 and 6 meters, depending on the snowpack. Thus, considering $L \sim 1\ m$ does not allow to cross out some terms, and justifies the failure of this strategy of full rescaling of the equations. It therefore seems necessary to solve the equations of motion fully numerically with due boundary conditions.

# Supplementary Material n°5 - Intrinsic limitations of the (Heierli, 2005) set of boundary conditions for sub-Rayleigh speeds

Both the *Heierli* and *generalised Heierli* solutions have a fundamental flaw related to the value of the curvature imposed at $x' = L$. The following development highlights the origin of the problem. We recall the equation of motion for the bending slab:

$$\left(1-\frac{c^2}{c_s^2}\right)\left(1-\frac{c^2}{c_p^2}\right)V^{(4)}+\frac{1}{\lambda^2}\frac{c^2}{c_s^2}V^{(2)}=\frac{1}{(\lambda h)^2}\Sigma_{eff}$$

It is a parametric differential equation in the velocity $c$, which has two singular values in the sub-Rayleigh regime: $c = c_s$ and $c = 0$ (plus one in the supershear regime, $c = c_p$). They show a very distinct boundary behaviour, depending on whether $c \to c_s$ or $c \to 0$. The regular solution ($\forall c \notin \{0, c_s\}$) should be able to connect to these two solutions for the corresponding velocity boundary values.

## 1. Static limit ($c = 0$)

The static limit has been extensively analysed by Rosendahl and Weißgraeber (2020a). In this case, only the fourth derivative remains. The equation of motion simplifies to:

$$V^{(4)}=\frac{1}{(\lambda h)^2}\Sigma_{eff}$$

whose general solution is a *fourth-order gravity-driven polynomial*:

$$V(x')=\frac{1}{24}\frac{\Sigma_{eff}}{\lambda^2 h^2}x'^4+Px'^3+Qx'^2+Rx'+S$$

with $P, Q, R, S$ four integration constants. Since the static length of the bending span $L_0$ is also unknown, five constants have to be determined to close the system: **five** boundary conditions have to be provided.

## 2. High speed limit ($c = c_s$)

When the velocity reaches that of transverse plane waves, only the second derivative remains and the equation simplifies to:

$$V^{(2)}=\frac{\Sigma_{eff}}{h^2}$$

whose general solution is a *gravity-driven parabola*, namely a parabola whose range $L$ is determined by the free fall time of the slab:

$$V(x')=\frac{1}{2}\frac{\Sigma_{eff}}{h^2}x'^2+T_1x'+T_2$$

where $T_1$, $T_2$ are integration constants. The length of the corresponding bending span $L_s$ is also unknown, so that the problem requires **three** boundary conditions to be fully determined.

## 3. Regular solution ($c \in ]0, c_s[$)

The solution of the regular problem has already been put forward in the framework of the *generalised Heierli model* (equation (21)). It can be written as

$$V(x') = A'r^4 \cos\left(\frac{x'}{r}\right) + B'r^3 \sin\left(\frac{x'}{r}\right) + \frac{1}{2}\frac{\Sigma_{eff}}{h^2}\frac{c_s^2}{c^2}\ x'^2 + C'x' + D'$$

This solution exhibits the two previous behaviours, with the gravitational term $(\frac{1}{2}\frac{\Sigma_{eff}}{h^2}\frac{c_s^2}{c^2}\ x'^2 + C'x' + D')$ expected to predominate at high speeds, and a harmonic term $(A'r^4cos(\frac{x'}{r}) + B'r^3sin(\frac{x'}{r}))$ allowing regularisation of the problem at low and medium speeds. To close the problem, **six** boundary conditions are required, four for the integration constants $A', B', C', D'$, one for the range $L$ and one for the speed $c$.

## 4. Low speed parametric assembly

In the static case, we choose to enforce the following four boundary conditions:

$$\begin{aligned} V(0) &= \frac{h_c}{h} \\ V(L) &= 0 \\ V'(0) &= 0 \\ V'(L) &= 0 \end{aligned}$$

and the solution (from Supplementary Material n°6) is

$$V(x', c = 0) = \frac{\Sigma_{eff}}{24h^2\lambda^2}x'^4 + \left(\frac{2h_c}{hL_0^3} - \frac{\Sigma_{eff}L_0}{12h^2\lambda^2}\right)x'^3 + \left(\frac{\Sigma_{eff}}{24h^2\lambda^2}L_0^2 - \frac{3h_c}{hL_0^2}\right)x'^2 + 0x' + \frac{h_c}{h} \tag{53}$$

An additional condition gives the static bending length $L_0$. For example, considering $V''(0) = 0$ gives:

$$\frac{\Sigma_{eff}}{24h^2\lambda^2}L_0^2 - \frac{3h_c}{hL_0^2} = 0 \Leftrightarrow L_0 = \sqrt[4]{\frac{72hh_c\lambda^2}{\Sigma_{eff}}}$$

The dynamic solution must now be connected to this static solution in its low-speed limit ($\forall x'$). As $r \to \infty$ when $c \to 0$, a limited expansion allows to write:

$$\begin{aligned} V(x', c \to 0) &= A'r^4 \cos\left(\frac{x'}{r}\right) + B'r^3 \sin\left(\frac{x'}{r}\right) + \frac{1}{2}\frac{\Sigma_{eff}}{h^2}\frac{r^2}{\lambda^2\Pi^2(c)}\ x'^2 + C'x' + D' \\ &\approx A'r^4\left(1 - \frac{x'^2}{2r^2} + \frac{x'^4}{24r^4}\right) + B'r^3\left(1 - \frac{x'^3}{6}\right) + \frac{1}{2}\frac{\Sigma_{eff}}{h^2}\frac{r^2}{\lambda^2}\ x'^2 + C'x' + D' \end{aligned}$$

$$V(x', c \to 0) \approx \frac{A'}{24}x'^4 - \frac{B'}{6}x'^3 + \frac{1}{2}\left(\frac{\Sigma_{eff}}{h^2\lambda^2}r^2 - A'r^2\right)x'^2 + (C' + B'r^2)x' + (D' + A'r^4) \tag{54}$$

The matching of (53) et (54) when $c \to 0$ imples:

$$\begin{aligned} L &\to \sqrt[4]{\frac{72hh_c\lambda^2}{\Sigma_{eff}}} \equiv L_0 \\ A' &\to \frac{\Sigma_{eff}}{\lambda^2h^2} \equiv A'_0 \\ B' &\to \frac{2h_c}{h} - \frac{\Sigma_{eff}L_0}{12h^2\lambda^2} = \frac{1}{3}A'_0L_0 \equiv B'_0 \\ C' &+ B'r^2 = 0 \end{aligned}$$

$$D' + A'r^4 = \frac{h_c}{h}$$

where the subscripts 0 denote the values of quantities that admit a finite limit when $c \to 0$.

Performing the dynamic assembly for $c \notin \{0, c_s\}$ with the same boundary conditions shows that the previous constraints on $A', B', C'$ and $D'$ are well verified, provided we assume that $L$ remains bounded. Now the static problem is fully determined, which implies in particular that $V^{(2)}(L_0)$ is enforced:

$$V^{(2)}(L_0) = \frac{A'_0 L_0^2}{6}$$

An additional condition is therefore added to the dynamic problem:

$$V^{(2)}(L, c \to 0) \to V^{(2)}(L_0) = \frac{A'_0 L_0^2}{6}$$

$$\Leftrightarrow A'r^2 \cos\left(\frac{L}{r}\right) + B'r^3 \sin\left(\frac{L}{r}\right) + \frac{\Sigma_{eff}}{h^2}\frac{c_s^2}{c^2} \to \frac{A'_0 L_0^2}{6}$$

However, the boundary condition $V^{(2)}(L) = \frac{\Sigma_{eff}}{h^2}\frac{c_s^2}{c^2}$ , used to obtain the *Heierli solution* and the *generalised Heierli solution*, does not meet the previous constraint. In fact, the form $\frac{\Sigma_{eff}}{h^2}\frac{c_s^2}{c^2}$ is only valid at the singular value $c = c_s$ where we recover the shear case. Setting it for any other speed is assuming that the fall at the anticrack tip is always gravitational (free fall), which is erroneous at low speeds where internal forces play an important role in regulating the acceleration. Paradoxically, thus, the *Heierli formula* seems to underestimate the velocity, whereas it is only valid in the high-speed limit.

## 5. High-speed parametric assembly

Since the singular value $c = c_s$ calls for only three boundary conditions to completely determine the shape and range of the bending, and being parabolic in shape, it is impossible for it to satisfy the double horizontal tangency imposed at $x' = 0$ and $x' = L$. Accordingly, no solution satisfying at least the five static boundary conditions can propagate at $c = c_s$. This is therefore an upper bound on the speed, which the system never reaches. Nevertheless, it is expected from the high-speed connection that

$$V^{(2)}(L, c \to c_s) \to V^{(2)}(L, c = c_s) = \frac{\Sigma_{eff}}{h^2} = \lambda^2 A'_0$$

which is the value reached by the slab during a free fall.

# Supplementary Material n°6 - Static solution of the bending of the slab

Let us perform the static assembly of the slab bending. The equation of motion is

$$V(x', c=0) = \frac{1}{24}\frac{\Sigma_{eff}}{\lambda^2 h^2}x'^4 + Px'^3 + Qx'^2 + Rx' + S$$

and the boundary conditions are:

$V(0) = S = 0$
$V'(0) = R = 0$
$V(L_0) = \frac{\Sigma_{eff}}{24\lambda^2 h^2}L_0^4 + PL_0^3 + QL_0^2 + \frac{h_c}{h} = 0$ (1)
$V'(L_0) = \frac{\Sigma_{eff}}{6\lambda^2 h^2}L_0^3 + 3PL_0^2 + 2QL_0 = 0$ (2)

Thus,

$L_0 \cdot (1) - 2 \cdot (2) \rightarrow \frac{\Sigma_{eff}}{12h^2\lambda^2}L_0^4 + PL_0^3 = \frac{2h_c}{h} \Leftrightarrow P = \frac{2h_c}{h} - \frac{\Sigma_{eff}L_0}{12h^2\lambda^2}$

$(2) \rightarrow Q = -\frac{1}{2}\left[3PL_0 + \frac{\Sigma_{eff}}{6\lambda^2 h^2}L_0^2\right] = -\frac{3h_c}{hL_0^2} + \frac{1}{24}\frac{\Sigma_{eff}}{h^2\lambda^2}L_0^2$

Hence

$$V(x', c=0) = \frac{\Sigma_{eff}}{24h^2\lambda^2}x'^4 + \left(\frac{2h_c}{hL_0^3} - \frac{\Sigma_{eff}L_0}{12h^2\lambda^2}\right)x'^3 + \left(\frac{\Sigma_{eff}}{24h^2\lambda^2}L_0^2 - \frac{3h_c}{hL_0^2}\right)x'^2 + \frac{h_c}{h}$$

## Supplementary Material n°7 - Dispersion curve from the dynamic solution of the bending of the slab

This assembly has already been performed in Section V of Supplementary Material n°2. We recall that

$$\gamma = \frac{L}{r}$$
$$r^2 = \lambda^2 \frac{c_s^2}{c^2} \Pi^2(c)$$
$$\Pi^2(c) = \left(1 - \frac{c^2}{c_s^2}\right)\left(1 - \frac{c^2}{c_p^2}\right)$$

The boundary conditions are written as follows:

$$A' r^4 + D' = \frac{h_c}{h} \tag{55}$$
$$A' r^4 \cos(\gamma) + B' r^3 \sin(\gamma) + \frac{1}{2}\frac{\Sigma_{eff}}{h^2}\frac{c_s^2}{c^2}\ L^2 + C' L + D' = 0 \tag{56}$$
$$B r^2 + \mathrm{C}' = 0 \tag{57}$$
$$-A' r^3 \sin(\gamma) + B' r^2 \cos(\gamma) + \frac{\Sigma_{eff}}{h^2}\frac{c_s^2}{c^2} L + C' = 0 \tag{58}$$
$$-A' r^2 + \frac{\Sigma_{eff}}{h^2}\frac{c_s^2}{c^2} = 0 \tag{59}$$
$$V^{(2)}(L) = -(A' r^2 \cos(\gamma) + B' r \sin(\gamma)) + \frac{\Sigma_{eff}}{h^2}\frac{c_s^2}{c^2} \tag{60}$$

so that

$$D' + A' r^4 = \frac{h_c}{h}$$
$$C' + B' r^2 = 0$$
$$A' = \frac{\Sigma_{eff}}{h^2}\frac{c_s^2}{c^2}\frac{1}{r^2} = \frac{\Sigma_{eff}}{h^2 \lambda^2}\frac{1}{\Pi^2} = \frac{A_0'}{\Pi^2} \underset{c\to 0}{\longrightarrow} A_0'$$

These equations allow the first three static conditions to be recovered:

(57)-(58) → $\quad (1-\cos(\gamma)) r^2 B' - r^3 \sin(\gamma)\, A' = \frac{\Sigma_{eff}}{h^2}\frac{c_s^2}{c^2} L = A' r^2 L$

$\Leftrightarrow (1-\cos(\gamma)) B' = r(\gamma - \sin(\gamma)) A'$

$\Leftrightarrow \frac{B'}{A'} = r\,\frac{\gamma - \sin(\gamma)}{1-\cos(\gamma)}$

The condition $B' \to \frac{A_0 L_0}{3} \equiv B_0'$ stems from the previous equation when $\gamma \underset{\substack{c\to 0\\ r\to\infty}}{\longrightarrow} 0$, which occurs if $L \underset{\substack{c\to 0\\ r\to\infty}}{\longrightarrow} L_0$.

(56) - $L \cdot$ (58) → $\quad A' r^4 \Big(\cos(\gamma) + L\frac{\sin(\gamma)}{r} - 1\Big) + B' r^3 \Big(\sin(\gamma) - L\frac{\cos(\gamma)}{r}\Big) - \frac{A' r^2}{2} L^2 = -\frac{h_c}{h}$

$$\Leftrightarrow L^2 = \frac{2h_c}{hA' r^2} + r^2(1 - \cos(\gamma) - \gamma\sin(\gamma)) + \frac{B'}{A'} r(\gamma\cos(\gamma) - \sin(\gamma))$$

and with (57)-(58),

$$L^2 = \frac{2hh_c}{\Sigma_{eff}}\frac{c^2}{c_s^2} + r^2(1-\cos(\gamma))\left(1 - \frac{\gamma\sin(\gamma)}{1-\cos(\gamma)} + \frac{\gamma\cos(\gamma) - \sin(\gamma)}{\gamma - \sin(\gamma)}\right) \tag{61}$$

## Supplementary Material n°8 - Relation between partial derivatives in steady state

In steady state, $V(x,t) = \bar{V}(x')$ where $x' = x - ct$. In particular,

$$\frac{\partial x'}{\partial x} = 1$$
$$\frac{\partial x'}{\partial t} = -c$$

Then,

$$\frac{\partial V}{\partial x} = \frac{\partial \bar{V}}{\partial x'}\frac{\partial x'}{\partial x} = \frac{d\bar{V}}{dx}\overbrace{\frac{\partial x'}{\partial x}}^{1} = V^{(1)}$$
$$\frac{\partial V}{\partial t} = \frac{\partial \bar{V}}{\partial x'}\frac{\partial x'}{\partial t} = \frac{d\bar{V}}{dx}\overbrace{\frac{\partial x'}{\partial t}}^{-c} = -cV^{(1)}$$
$$\frac{\partial^2 V}{\partial x^2} = \frac{\partial}{\partial x}\left(\frac{\partial V}{\partial x}\right) = \frac{\partial}{\partial x'}\left(\frac{\partial V}{\partial x}\right)\overbrace{\frac{\partial x'}{\partial x}}^{1} = \frac{\partial}{\partial x'}(\bar{V}^{(1)}) = \frac{d\bar{V}^{(1)}}{dx'} = V^{(2)}$$
$$\frac{\partial^2 V}{\partial t^2} = \frac{\partial}{\partial t}\left(\frac{\partial V}{\partial t}\right) = \frac{\partial}{\partial x'}\left(\frac{\partial V}{\partial t}\right)\overbrace{\frac{\partial x'}{\partial t}}^{-c} = -c\frac{\partial}{\partial x'}(-c\bar{V}^{(1)}) = c^2\frac{d\bar{V}^{(1)}}{dx'} = c^2V^{(2)}$$

from which the following equations arise:

$$\frac{d\bar{V}}{\partial x'} = \frac{\partial V}{\partial x} = -\frac{1}{c}\frac{\partial V}{\partial t}$$

$$\frac{d^2\bar{V}}{\partial x'^2} = \frac{\partial^2 V}{\partial x^2} = \frac{1}{c^2}\frac{\partial^2 V}{\partial t^2}$$

## Supplementary references